\documentclass[%
reprint,
onecolumn,
amsmath,
amssymb,
aip,
pof,
]{revtex4-1}

\usepackage{graphicx}
\usepackage{dcolumn}
\usepackage{bm}
\usepackage{epsfig}
\usepackage{tabularx}
\usepackage{color}
\usepackage{verbatim}

\begin{document}
\title{Modeling realistic multiphase flows using a non-orthogonal multiple-relaxation-time lattice Boltzmann method}

\author{Linlin Fei}
\affiliation{Center for Combustion Energy; Key Laboratory for Thermal Science and Power Engineering of Ministry of Education, Department of Energy and Power Engineering, Tsinghua University, Beijing 100084, China.}
\affiliation{
	Istituto per le Applicazioni del Calcolo, Consiglio Nazionale delle Ricerche, Via dei Taurini 19, 00185 Rome, Italy}

\author{Jingyu Du}
\affiliation{
	Institute of Nuclear and New Energy Technology, Tsinghua University, Key Laboratory for Advanced Reactor Engineering and Safety of the Ministry of Education, Beijing 100084, China}

\author{Kai H. Luo}
\email{K.Luo@ucl.ac.uk}
\affiliation{Center for Combustion Energy; Key Laboratory for Thermal Science and Power Engineering of Ministry of Education, Department of Energy and Power Engineering, Tsinghua University, Beijing 100084, China.}
\affiliation{Department of Mechanical Engineering, University College London, Torrington Place, London WC1E 7JE, UK}

\author{Sauro Succi}
\affiliation{
	Istituto per le Applicazioni del Calcolo, Consiglio Nazionale delle Ricerche, Via dei Taurini 19, 00185 Rome, Italy}
\affiliation{Center for Life Nano Science at La Sapienza, Istituto Italiano di Tecnologia, 295 Viale Regina Elena, I-00161 Roma, Italy}
\affiliation{Harvard Institute for Applied Computational Science, Cambridge, Massachusetts 02138, USA}

\author{Marco Lauricella}
\affiliation{
	Istituto per le Applicazioni del Calcolo, Consiglio Nazionale delle Ricerche, Via dei Taurini 19, 00185 Rome, Italy}

\author{Andrea Montessori}
\affiliation{
	Istituto per le Applicazioni del Calcolo, Consiglio Nazionale delle Ricerche, Via dei Taurini 19, 00185 Rome, Italy}

\author{Qian Wang}
\affiliation{School of Mechanical Engineering, Shanghai Jiao Tong University, Shanghai 200240, China}
\date{\today}

\begin{abstract}
In this paper, we develop a  three-dimensional multiple-relaxation-time lattice Boltzmann method (MRT-LBM) based on a
set of non-orthogonal basis vectors.
Compared with the classical MRT-LBM based on a set of orthogonal basis vectors, the present non-orthogonal MRT-LBM simplifies the transformation between the discrete velocity space and the moment space, and exhibits better portability across different lattices. The proposed method is then extended to multiphase flows at large density ratio with tunable surface tension, and its numerical stability and accuracy are well 
demonstrated by some benchmark cases. Using the proposed method, 
a practical case of a fuel droplet impacting on a dry surface at high Reynolds and Weber numbers is simulated and the evolution of the spreading film diameter agrees well with the experimental data. Furthermore, another realistic case of a droplet impacting on a super-hydrophobic wall with a cylindrical obstacle is reproduced, which confirms the experimental finding of Liu \textit{et al.} [``Symmetry breaking in drop bouncing on curved surfaces," Nature communications 6, 10034 (2015)] that the contact time is minimized when the cylinder radius is comparable with the droplet cylinder.

\end{abstract}
\maketitle


\section{Introduction}
Interfaces between different phases and/or components are ubiquitous in multiphase flows and energy applications, such as rain dynamics, plant spraying, water boiling, gas turbine blade cooling, to name but a few \cite{richard2002surface,li2016lattice}. A deeper understanding of the fundamental physics of such complex interfaces is of great importance in many natural and industrial processes. The dynamics of the interfaces is difficult to investigate because typical interfaces are extremely thin, complex in shape and deforming at short time scales. In addition, the  density ratio, Weber and Reynolds numbers involved in many practical multiphase flows, such as binary droplets collisions and melt-jet breakup, are usually very high, which further increases the complexity of the phenomena involved. Therefore, development of robust and accurate computational methods to capture the complex interfacial phenomena is crucial in the study of multiphase flows.
 
During  the last three decades, the mesoscopic lattice Boltzmann method (LBM), based on the kinetic theory, has become an increasingly important method for numerical simulations of multiphase flows, mainly on account of its meso-scale features, easy implementation and computational efficiency \cite{shan1993lattice,yuan2006equations,gan2011lattice,li2012forcing,lycett2014binary,monaco2014numerical,succi2015lattice,li2015lattice,lycett2015improved,lycett2016cascaded,mazloomi2016simulation,fu2017theoretical,saito2017lattice,saito2018color,qin2018entropic,Cuiyutong2019,chenguoqing2019}. Generally, the existing multiphase LB models can be classified into four categories: the color-gradient model \cite{gunstensen1991lattice,grunau1993lattice}, the pseudopotential model \cite{shan1993lattice,shan1994simulation}, the free-energy model \cite{swift1995lattice,swift1996lattice} and the mean-field model \cite{he1999lattice}.
Among them, the pseudopotential model is considered in the present work due to its simplicity and computational efficiency. In this model, the interactions among populations of molecules are modeled by a density-dependent pseudopotential. Through interactions among the particles on the nearest-neighboring sites,  phase separation and breakup and/or merging of phase interfaces can be achieved automatically. For further details about the multiphase LB models, interested readers are directed to some comprehensive review papers \cite{li2016lattice,chen2014critical,liu2016multiphase}.

In the LBM framework, the fluid is usually represented by populations of fictitious
particles colliding locally and streaming to adjacent nodes  along the links of a regular lattice. The macroscopic variables are obtained through a set of rules based on the calculated particle distribution functions (DFs).  In particular, the simplest scheme to execute the ``colliding" (or collision)  step is to relax all the distribution functions (DFs) to their local equilibria at an identical rate, known as the single-relaxation-time (SRT) scheme \cite{qian1992lattice}. However, SRT-LBM usually suffers numerical instability for flows with even moderate Reynolds number. Compared with SRT scheme, the multi-relaxation-time (MRT) scheme, originally formulated in \cite{higuera1989lattice,d1994generalized} and later extended in \cite{lallemand2000theory,d2002multiple}, is able to enhance the stability by carefully separating the time scales among the kinetic modes. To enhance numerical stability of LBM, some modified approaches within the SRT framework have also been proposed, such as  the entropic LBM \cite{karlin1999perfect,ansumali2000stabilization} and regularized LBM \cite{zhang2006efficient,latt2006lattice}. In addition, the cascaded lattice Boltzmann method (CLBM), which employs moments in a co-moving frame in contrast to the stationary moments in MRT, has also been shown to improve numerical stability significantly compared with the classical SRT-LBM \cite{geier2006cascaded,lycett2014binary,lycett2016cascaded,de2017nonorthogonal,fei2018modeling,saito2018color}.

The present work focuses on the MRT-LBM, in which the collision step is carried out in a (raw) moment space via a transformation matrix  $ {\bf{M}} $, where different moments can be relaxed independently. The  post-collision moments are then transformed back via
$ {{\bf{M}}^{{\rm{ - }}1}} $ and the streaming step is implemented in the discrete velocity space as usual. Usually, the Gram-Schmidt procedure is adopted to construct an orthogonal transformation matrix \cite{lallemand2000theory,d2002multiple,suga2015d3q27,saito2017lattice}, which means that the basis vectors for the moments are orthogonal to one another. 
It is known that the widely used orthogonal MRT-LBM is more complex and computationally expensive than SRT-LBM, especially for three-dimensional problems. To the best of our knowledge, orthogonality is not a necessary condition for stability. As an early attempt, Lycett-Brown and Luo \cite{lycett2014multiphase} showed that an MRT-LBM based on a non-orthogonal basis vector set enhances the numerical stability compared with the SRT-LBM. The corresponding non-orthogonal MRT-LBM has been 
extended to simulate incompressible thermal flows by Liu \textit{et al.} \cite{liu2016non}.  In addition, it was shown by Li \textit{et al.} \cite{li_non_mrt} that a non-orthogonal MRT-LBM can retain the numerical accuracy while simplifying the implementation of its orthogonal counterpart. In parallel, the CLBM \cite{geier2006cascaded}, which can be viewed as a non-orthogonal MRT-LBM in the co-moving frame, has been shown to possess very good numerical stability for high Rayleigh number thermal flows \cite{fei2018three}, as well as high Reynolds and Weber numbers multiphase flows \cite{lycett2014binary,saito2018color,lycett2016cascaded}. Recently, an improved three-dimensional (3D) CLBM has been proposed by Fei \textit{et al.} \cite{fei2018modeling}, where an improved set of non-orthogonal basis vectors was employed and a generalized multiple-relaxation-time (GMRT) scheme \cite{fei2018modeling,fei2017consistent} was adopted to cast MRT-LBM and CLBM into a unified framework. Within the GRMT framework, the CLBM can reduce to a non-orthogonal MRT-LBM when the shift matrix is a unit matrix, where the shift matrix is defined to shift
 (raw) moments of the DFs to the corresponding central moments. 

In this work, we first give a 
theoretical analysis to construct a generalized non-orthogonal MRT-LBM based on the basis vector set proposed in  Ref. \cite{fei2018three}. Coupled with the pseudopotential multiphase model, the proposed non-orthogonal MRT-LBM is extended to simulate
multiphase flows with large density ratio and tunable surface tension, which is then verified by some benchmark cases. Finally, we provide simulations of two practical and challenging problems using our proposed non-orthogonal MRT-LBM, to highlight its capability for simulating realistic multiphase flows.

\section{Non-orthogonal MRT-LBM for multiphase flows}\label{sec.2}
The theoretical derivation of the non-orthogonal MRT-LBM is given in this section. Firstly, the MRT framework is introduced briefly. Then, the choice of the non-orthogonal basis vector set is presented. In the end, the pseudopotential model is incorporated into the present method to simulate multiphase flows.
\subsection{MRT framework}\label{sec.2a}
In this section, the MRT-LBM framework is introduced based on the standard D3Q27 discrete velocity model (DVM). However, it should be noted that the procedures shown in this work are not limited to the specified DVM, and can be extended to other DVMs readily. The lattice speed $ c = \Delta x = \Delta t = 1 $ and the lattice sound speed  ${c_s} = 1/\sqrt 3 $ are adopted, in which  $ \Delta x  $ and   $ \Delta t  $ are the lattice spacing and time step, respectively. The discrete velocities $ {{\bf{e}}_i} = [\left| {{e_{ix}}} \right\rangle ,\left| {{e_{iy}}} \right\rangle ,\left| {{e_{iz}}} \right\rangle ]  $ are defined as
\begin{equation}\label{e1}
\begin{array}{l}
\left| {{e_{ix}}} \right\rangle  = {[0,1, - 1,0,0,0,0,1, - 1,1, - 1,1, - 1,1, - 1,0,0,0,0,1, - 1,1, - 1,1, - 1,1, - 1]^ \top }, \\ 
\left| {{e_{iy}}} \right\rangle  = {[0,0,0,1, - 1,0,0,1,1, - 1, - 1,0,0,0,0,1, - 1,1, - 1,1,1, - 1, - 1,1,1, - 1, - 1]^ \top }, \\ 
\left| {{e_{i{\rm{z}}}}} \right\rangle  = {[0,0,0,0,0,1, - 1,0,0,0,0,1,1, - 1, - 1,1,1, - 1, - 1,1,1,1,1, - 1, - 1, - 1, - 1]^ \top }. \\ 
\end{array}
\end{equation}
where $ i = 0,1,...,26 $, ${\left|  \cdot  \right\rangle }$ denotes a 27-dimensional column vector, and the superscript $ \top $ denotes the transposition.  

To execute the collision step in the moment space, we first define moments of
the discrete distribution function (DFs) $ {f_i} $,	
\begin{equation}\label{e2}
{k_{mnp}} = \left\langle {{f_i}|e_{ix}^me_{iy}^ne_{iz}^p} \right\rangle,
\end{equation}
where $ m $, $ n $, and $ p $ are integers.  The equilibrium moments $ k_{mnp}^{eq} $  are defined analogously by replacing $ {f_i} $  with the discrete equilibrium distribution functions (EDFs) $ f_i^{eq} $. To construct an MRT-LBM, an appropriate moment set vector $ {\bf{m}} $ is needed,
\begin{equation}\label{e3}
{\bf{m}} = {[{m_0},{m_1},...,{m_{26}}]^{\rm T}}
\end{equation}
where the elements in   $ {\bf{m}} $  are combinations of $ {k_{mnp}} $. The transformation from the discrete velocity space to the moment space can be performed through a transformation matrix $ {\bf{M}} $  by  $ {\bf{m}} = {\bf{Mf}} $. The explicit expression for  $ {\bf{M}} $ depends on the raw moment set in Eq. (\ref{e3}) , which will be discussed in the next subsection.

A general collision step in MRT-LBM can be written as \cite{li2010improved},
\begin{equation}\label{e4}
f_i^*({\bf{x}},t) = {f_i}({\bf{x}},t) - {\Lambda _{i,k}}[{f_k} - f_k^{eq}]{|_{({\bf{x}},t)}} + \frac{{\Delta t}}{2}[{\bar F_i}({\bf{x}},t) + {\bar F_i}({\bf{x}} + {{\bf{e}}_i}\Delta t,t + \Delta t)],
\end{equation}
where $ {\bf{x}} $  is the spatial position, $ t $ is time,  $ {\bar F_i} $ are the forcing terms in the discrete velocity space, and $ {\Lambda _{i,k}} = {({{\bf{M}}^{{\bf{ - 1}}}}{\bf{SM}})_{i,k}} $ is the collision operator, in which $ {\bf{S}} $  is a diagonal relaxation matrix. The EDFs $ f_i^{eq}  $ are often given by a low-Mach truncation form,
\begin{equation}\label{e5}
f_i^{eq} = \rho \omega ({\left| {{{\bf{e}}_i}} \right|^2})\left[ {1 + \frac{{{{\bf{e}}_i}\cdot{\bf{u}}}}{{c_s^2}} + \frac{{{\bf{uu}}:({{\bf{e}}_i}{{\bf{e}}_i} - c_s^2)}}{{c_s^4}}} \right]
\end{equation}
where $\rho $ is the fluid density,  $ {\bf{u}} = [{u_x},{u_y},{u_z}] $ is the fluid velocity, and  the weights are $\omega (0) = 8/27 $, $\omega (1) = 2/27$, $\omega (2) = 1/54$ and $\omega (3) = 1/216$. According to the analysis by Guo \textit{et al}. \cite{guo2002discrete}, the forcing terms are defined as,
\begin{equation}\label{e6}
{{\bar F}_i}{\rm{ = }}\omega ({\left| {{{\bf{e}}_i}} \right|^2})\left[ {\frac{{{{\bf{e}}_i} - {\bf{u}}}}{{c_s^2}} + \frac{{({\bf{u}}\cdot{{\bf{e}}_i}){{\bf{e}}_i}}}{{c_s^4}}} \right]\cdot{\bf{F}},
\end{equation}
where  $ {\bf{F}} = [{F_x},{F_y},{F_z}] $ is the total force exerted on the fluid system.

To remove the implicit implementation, Eq. (\ref{e4}) can be modified as,
\begin{equation}\label{e7}
{\bar f_i}({\bf{x}} + {{\bf{e}}_i}\Delta t,t + \Delta t) = {\bar f_i}({\bf{x}},t) - {\Lambda _{i,k}}[{\bar f_k} - f_k^{eq}]{|_{({\bf{x}},t)}} + ({\bf{I}} - \frac{{{\Lambda _{i,k}}}}{2}){\bar F_i}({\bf{x}},t)\Delta t,
\end{equation}
where $ {\bar f_i} = {f_i} - \Delta t{\bar F_i}/2 $ and $ {\bf{I}} $ is the unit matrix. Multiplying Eq. (\ref{e7}) by the transformation matrix  $ {\bf{M}} $, the collision step in the moment space can be rewritten as
\begin{equation}\label{e8}
{{\bf{\bar m}}^*} = {\bf{\bar m}} - {\bf{S}}({\bf{\bar m}} - {{\bf{\bar m}}^{eq}}) + ({\bf{I}} - \frac{{\bf{S}}}{2})\Delta t{\bf{\tilde F}},
\end{equation}
where $ {\bf{\bar m}} = {\bf{M\bar f}} $, $ {{\bf{m}}^{eq}} = {\bf{M}}{{\bf{f}}^{eq}} $ and $ {\bf{\tilde F}} = {\bf{M\bar F}} $.

After the collision step, post-collision discrete DFs can be reconstructed by   ${{\bf{\bar f}}^{\rm{*}}} = {{\bf{M}}^{ - 1}}{{\bf{\bar m}}^*}
 $. In the streaming step, the post-collision discrete DFs in space  $ {\bf{x}} $ stream to their neighbors  $ ({\bf{x}} + {{\bf{e}}_i}\Delta t) $ along the characteristic lines as usual,
\begin{equation}\label{e9}
{\bar f_i}({\bf{x}} + {{\bf{e}}_i}\Delta t,t + \Delta t) = \bar f_i^*({\bf{x}},t)
\end{equation}
The hydrodynamic variables are updated by
\begin{equation}
\rho  = \sum\limits_i {{{\bar f}_i},} ~~~{\rm{   }}\rho {\bf{u}} = \sum\limits_i {{{\bar f}_i}} {{\bf{e}}_i} + \frac{{\Delta t{\bf{F}}}}{2}.
\end{equation}
\subsection{Non-orthogonal basis vector set}\label{sec.2b}
In this work, we adopt a  moment set $  {\bf{m}} = {[{m_0},{m_1},...,{m_{26}}]^{\rm T}} $  with the following 27 moment elements (in the ascending order of $ m + n + p $ ),
\begin{equation}\label{e11}
\begin{array}{l}
{\bf{m}} = [{k_{000}},{k_{100}},{k_{010}},{k_{001}},{k_{110}},{k_{101}},{k_{011}},{k_{200}} + {k_{020}} + {k_{002}},{k_{200}} - {k_{020}},{k_{200}} - {k_{002}},{k_{120}}, \\ 
{\rm{       }}{k_{102}},{k_{210}},{k_{201}},{k_{012}},{k_{021}},{k_{111}},{k_{220}},{k_{202}},{k_{022}},{k_{211}},{k_{121}},{k_{112}},{k_{122}},{k_{212}},{k_{221}},{k_{222}}{]^{\rm T}} \\ 
\end{array}
\end{equation}
where the elements $ {m_0}$ , ${m_{1 - 3}}$, and ${m_{4 - 9}}$ are related to the fluid density, momentum, and viscous stress tensor, respectively, while the remaining elements are higher-order moments which do not affect the consistency at the Navier-Stokes level. It should be pointed out that  the moments are chosen based on two criteria: (i) the basis vectors for the moments are linearly independent (but not necessarily orthogonal to one another); (ii) the calculation of each moment is as simple as possible. Generally, the high-order elements are related to some kinetic moments, such as energy flux and square of kinetic energy, but the relations are not defined exactly. The above moment set in Eq. (\ref{e11}) was originally adopted in our cascaded LBM to improve the implementation \cite{fei2018three}, where the mixed second-order moments $({k_{200}} + {k_{020}} + {k_{002}},{k_{200}} - {k_{020}},{k_{200}} - {k_{002}})$ were implicit and the relaxation matrix was slightly modified  to  simplify the map between the (raw) moment space and central moment space. In the present paper, the relaxation matrix ${\bf{S}}$ is a diagonal matrix,
\begin{equation}\label{e12}
{\bf{S}} = diag({s_0},{s_1},{s_1},{s_1},{s_2},{s_2},{s_2},{s_{2b}},{s_2},{s_2},{s_3},{s_3},{s_3},{s_3},{s_3},{s_3},{s_{3b}},{s_4},{s_4},{s_4},{s_{4b}},{s_{4b}},{s_{4b}},{s_5},{s_5},{s_5},{s_6}),
\end{equation}
where the elements are the relaxation rates for different moments. The kinematic and bulk viscosities are related to the relaxation rates for the second-order moments by $ \nu  = (1/{s_2} - 0.5)c_s^2\Delta t $  and $ \xi  = 2/3(1/{s_{2b}} - 0.5)c_s^2\Delta t $, respectively. Here, we use ${s_0} = {s_1} = 1.0$, ${s_{2b}} = {s_{3b}} = 0.6$ and the others are set to be 1.2.

The transformation matrix $ {\bf{M}} $  can be obtained explicitly according to Eq. (\ref{e2}) and Eq. (\ref{e11}),

\begin{equation}\label{matrix}
\setlength{\arraycolsep}{1.2pt}
{\bf{M} = }\left[ 
\begin{array}{l r r r r r r r r r r r r r r r r r r r r r r r r r r}
1 &1 &1&1&1&1&1&1&1&1 &1 &1&1&1&1&1&1&1&1&1&1&1&1&1&1&1&1\\

0 &1 & -1&0&0&0&0&1&-1&1&-1&1&-1&1&-1&0&0&0&0&1&-1&1&-1&1&-1&1&-1\\
0 &0 &0&1&-1&0&0&1&1&-1 &-1 &0&0&0&0&1&-1&1&-1&1&1&-1&-1&1&1&-1&-1\\
0 &0 &0&0&0&1&-1&0&0&0&0 &1&1&-1&-1&1&1&-1&-1&1&1&1&1&-1&-1&-1&-1\\

0 &0 &0&0&0&0&0&1&-1&-1&1&0&0&0&0&0&0&0&0&1&-1&-1&1&1&-1&-1&1\\
0 &0 &0&0&0&0&0&0&0&0&0&1&-1&-1&1&0&0&0&0&1&-1&1&-1&-1&1&-1&1\\
0 &0 &0&0&0&0&0&0&0&0&0&0&0&0&0&1&-1&-1&1&1&1&-1&-1&-1&-1&1&1\\
0 &1 &1&1&1&1&1&2&2&2 &2 &2&2&2&2&2&2&2&2&3&3&3&3&3&3&3&3\\
0 &1 &1&-1&-1&0&0&0&0&0 &0 &1&1&1&1&-1&-1&-1&-1&0&0&0&0&0&0&0&0\\
0 &1 &1&0&0&-1&-1&1&1&1 &1 &0&0&0&0&-1&-1&-1&-1&0&0&0&0&0&0&0&0\\

0 &0 &0&0&0&0&0&1&-1&1 &-1 &0&0&0&0&0&0&0&0&1&-1&1&-1&1&-1&1&-1\\
0 &0 &0&0&0&0&0&0&0&0 &0 &1&-1&1 &-1&0&0&0&0&1&-1&1&-1&1&-1&1&-1\\
0 &0 &0&0&0&0&0&1&1&-1 &-1 &0&0&0&0&0&0&0&0&1&1&-1&-1&1&1&-1&-1\\
0 &0 &0&0&0&0&0&0&0&0 &0 &1&1&-1&-1&0&0&0&0&1&1&1&1&-1&-1&-1&-1\\
0 &0 &0&0&0&0&0&0&0&0 &0 &0&0&0&0&1&-1&1&-1&1&1&-1&-1&1&1&-1&-1\\
0 &0 &0&0&0&0&0&0&0&0 &0 &0&0&0&0&1&1&-1&-1&1&1&1&1&-1&-1&-1&-1\\

0 &0 &0&0&0&0&0&0&0&0 &0 &0&0&0&0&0&0&0&0&1&-1&-1&1&-1&1&1&-1\\

0 &0 &0&0&0&0&0&1&1&1 &1 &0&0&0&0&0&0&0&0&1&1&1&1&1&1&1&1\\
0 &0 &0&0&0&0&0&0&0&0&0 &1&1&1&1&0&0&0&0&1&1&1&1&1&1&1&1\\
0 &0 &0&0&0&0&0&0&0&0&0 &0&0&0&0&1&1&1&1&1&1&1&1&1&1&1&1\\
0 &0 &0&0&0&0&0&0&0&0&0 &0&0&0&0&0&0&0&0&1&1&-1&-1&-1&-1&1&1\\
0 &0 &0&0&0&0&0&0&0&0&0 &0&0&0&0&0&0&0&0&1&-1&1&-1&-1&1&-1&1\\
0 &0 &0&0&0&0&0&0&0&0&0 &0&0&0&0&0&0&0&0&1&-1&-1&1&1&-1&-1&1\\

0 &0 &0&0&0&0&0&0&0&0&0 &0&0&0&0&0&0&0&0&1&-1&1&-1&1&-1&1&-1\\
0 &0 &0&0&0&0&0&0&0&0&0 &0&0&0&0&0&0&0&0&1&1&-1&-1&1&1&-1&-1\\
0 &0 &0&0&0&0&0&0&0&0&0 &0&0&0&0&0&0&0&0&1&1&1&1&-1&-1&-1&-1\\

0 &0 &0&0&0&0&0&0&0&0&0 &0&0&0&0&0&0&0&0&1&1&1&1&1&1&1&1\\
\end{array} \right].
\end{equation}
The equilibrium raw moment vector ${{\bf{m}}^{eq}} = [m_0^{eq},m_1^{eq},...,m_{26}^{eq}]$ finally reads
\begin{equation}\label{e13}
\begin{array}{l}
{{\bf{m}}^{eq}} = [\rho ,\rho {u_x},\rho {u_y},\rho {u_z},\rho {u_x}{u_y},\rho {u_x}{u_z},\rho {u_y}{u_z},\rho (1 + {{\bf{u}}^2}),\rho (u_x^2 - u_y^2),\rho (u_x^2 - u_z^2), \\ 
\rho c_s^2{u_x},\rho c_s^2{u_x},\rho c_s^2{u_y},\rho c_s^2{u_z},\rho c_s^2{u_y},\rho c_s^2{u_z},0,\rho c_s^2(c_s^2 + u_x^2 + u_y^2),\rho c_s^2(c_s^2 + u_x^2 + u_z^2), \\ 
\rho c_s^2(c_s^2 + u_y^2 + u_z^2),\rho c_s^2{u_y}{u_z},\rho c_s^2{u_x}{u_z},\rho c_s^2{u_x}{u_y},\rho c_s^4{u_x},\rho c_s^4{u_y},\rho c_s^4{u_z},\rho c_s^4{{\bf{u}}^2} + \rho c_s^6{]^{\rm T}} \\ 
\end{array}
\end{equation}
and the forcing term vector in the moment space ${\bf{\tilde F}} = [{\tilde F_0},{\tilde F_1},...,{\tilde F_{26}}]$,
\begin{equation}\label{e14}
\begin{array}{l}
{\bf{\tilde F}} = [0,{F_x},{F_y},{F_z},{F_x}{u_y} + {F_y}{u_x},{F_x}{u_z} + {F_z}{u_x},{F_y}{u_z} + {F_z}{u_y},2{\bf{F}}\cdot{\bf{u}},2({F_x}{u_x} - {F_y}{u_y}),2({F_x}{u_x} - {F_{\rm{z}}}{u_z}), \\ 
{F_x}c_s^2,{F_x}c_s^2,{F_y}c_s^2,{F_z}c_s^2,{F_y}c_s^2,{F_z}c_s^2,0,2c_s^2({F_x}{u_x} + {F_y}{u_y}),2c_s^2({F_x}{u_x} + {F_z}{u_z}),2c_s^2({F_y}{u_y} + {F_z}{u_z}), \\ 
c_s^2({F_y}{u_z} + {F_z}{u_y}),c_s^2({F_x}{u_z} + {F_z}{u_x}),c_s^2({F_x}{u_y} + {F_y}{u_x}),c_s^4{F_x},c_s^4{F_y},c_s^4{F_z},2c_s^4{\bf{F}}\cdot{\bf{u}}{]^{\rm T}} \\ 
\end{array}
\end{equation}

It can be found that the transformation matrix ${\bf{M}}$ in Eq. (\ref{matrix}) is non-orthogonal. Through the Chapman-Enskog analysis (see Appendix \ref{sec.5a}), the proposed non-orthogonal MRT-LBM can recover the Navier-Stokes equations in the low Mach number limit. When all the relaxation parameters in the matrix  ${\bf{S}}$  are set equal, the present non-orthogonal MRT-LBM reduces to the SRT-LBM. 
Compared with the orthogonal MRT-LBM used in \cite{d2002multiple,suga2015d3q27}, the numbers of non-zeros in the present non-orthogonal ${\bf{M}}$  and its inverse matrix ${{\bf{M}}^{ - 1}}$
are much smaller (see in Table \ref{TABCOMPARE}), which indicates the implementation is simplified and the computational efficiency is enhanced.
Quantitatively, the non-orthogonal MRT-LBM requires approximately 25\% and 15\% less computational time for the D3Q27 model and D3Q19 model \cite{fei2018three,li_non_mrt}, respectively. Moreover, a non-orthogonal D3Q19 MRT-LBM can be extracted from the D3Q27 model directly (see Appendix
\ref{sec.5b}, all the elements in ${\bf{m}}$, ${{\bf{m}}^{eq}}$, ${\bf{s}}$, ${\bf{\tilde F}}$, ${\bf{M}}$, and ${{\bf{M}}^{ - 1}}$ for the D3Q19 model can be extracted from the D3Q27 model directly), which means that the non-orthogonal MRT-LBM exhibits very good portability across lattices. The comparison between the non-orthogonal and orthogonal MRT-LBMs on the D3Q19 lattice is also shown in Table \ref{TABCOMPARE}. Interested readers are kindly directed to the Supplementary Material for the explicit expressions of
$ {\bf{M}} $ and $ {{\bf{M}}^{ - 1}} $.

\begin{table*}
	\renewcommand\arraystretch{1.3}
	\caption{\label{TABCOMPARE}
		Non-zero numbers in $ {\bf{M}} $ and $ {{\bf{M}}^{{\rm{ - }}1}} $ for orthogonal and non-orthognal MRT models.
	}
	\centering
	\begin{tabular}{lcccc}
		\toprule
&\multicolumn{2}{c}{Orthogonal \cite{d2002multiple,suga2015d3q27}} &\multicolumn{2}{c}{Non-orthogonal}  \\ 
		\cline{2-3} \cline{4-5}			
		Models
		& D3Q27 & D3Q19 & D3Q27& D3Q19
		\\ 
		\hline			
		$ {\bf{M}} $&416&213&339&139\\	
		$ {{\bf{M}}^{{\rm{ - }}1}} $&416&213&226&101\\
		\toprule		
	\end{tabular}
\end{table*}
\subsection{Multiphase model}\label{sec.2c}
To extend the above mentioned non-orthogonal MRT-LBM to multiphase flows, the pseudopotential model \cite{shan1993lattice,shan1994simulation} is considered in the present work. It may be noted that the present non-orthogonal MRT-LBM can be also coupled with other multiphase models in a similar way. In the pseudopotential model, the interactions among molecules clusters are modeled by a pseudo-interaction force among fictitious particles,
\begin{equation}\label{e15}
{{\bf{F}}_{{\rm{int}}}} =  - G\psi ({\bf{x}})\sum\limits_i {w({{\left| {{{\bf{e}}_i}} \right|}^2})} \psi ({\bf{x}} + {{\bf{e}}_i}\Delta t){{\bf{e}}_i}
\end{equation}
where \textit{G} is the interaction strength, $ \psi $   is a density-dependent pseudopotential, and the normalized weights are $w({\left| {{{\bf{e}}_i}} \right|^2}) = \omega ({\left| {{{\bf{e}}_i}} \right|^2})/c_s^2$
. According to the Chapman-Enskog analysis, the bulk pressure reads,
\begin{equation}\label{e16}
p = \rho c_s^2 + \frac{{G{c^2}}}{2}{\psi ^2}
\end{equation}

In order to incorporate different equations of state consistently and achieve large density ratio, the square-root-form pseudopotential \cite{yuan2006equations} $ \psi  = \sqrt {2({p_{EOS}} - \rho c_s^2)/G{c^2}} $  is used in this work, where $  {p_{EOS}} $ is given by the adopted equation of state. Furthermore, some terms in $ {\bf{\tilde F}} $ need to be slightly modified for the sake of thermodynamic consistency and tunable surface tension \cite{li2012forcing,li2013lattice,li2013achieving}, and the details are provided in 
Appendix \ref{sec.5a2}.

\section{Numerical verification}\label{sec.3}
\subsection{Realization of large density ratio}\label{sec.3a}
First, we consider the verification of the liquid and vapor coexistence densities at large density ratios. To achieve large density ratios, different equations of state can be incorporated into the pseudopotential model, such as the Carnahan-Starling (C-S) equation, Peng-Robinson (P-R) equation and the piecewise linear equation of state \cite{li2013lattice,xu2015three,gong2017thermal,colosqui2012mesoscopic}. In this paper, we use the piecewise equation of state, which is given by \cite{colosqui2012mesoscopic},
\begin{equation}\label{e19}
p(\rho ) = \left\{ \begin{array}{l}
\rho {\theta _v},~~{\rm{    }}\rho  \le {\rho _1} \\ 
{\rho _1}{\theta _v} + (\rho  - {\rho _1}){\theta _m},~~{\rm{  }}{\rho _1} \le \rho  \le {\rho _2} \\ 
{\rho _1}{\theta _v} + ({\rho _2} - {\rho _1}){\theta _m} + (\rho  - {\rho _2}){\theta _l},~~{\rm{  }}\rho  > {\rho _2} \\ 
\end{array} \right.
\end{equation}
where  $ {\theta _v} = {(\partial p/\partial \rho )_v} > 0 $
,  $ {\theta _l} = {(\partial p/\partial \rho )_l} > 0 $
, and  $ {\theta _m} = {(\partial p/\partial \rho )_m} < 0 $
 are the slopes of $ p(\rho ) $  in the vapor-phase region, the liquid-phase region and the mechanically unstable region, respectively. In addition, $ \sqrt {{\theta _v}} $ and  $ \sqrt {{\theta _l}} $ can be regarded as the sound speed in the vapor and liquid phases, respectively. The unknown  $ {\rho _1} $ and $ {\rho _2} $,  defining the spinodal points are obtained by solving the following two equations, which is related to the mechanical and chemical equilibrium conditions,
 \begin{equation}\label{e20}
 \begin{array}{l}
 \int_{{\rho _v}}^{{\rho _l}} {({\rho _1} - {\rho _v}){\theta _v} + ({\rho _2} - {\rho _1})} {\theta _m} + ({\rho _l} - {\rho _2}){\theta _l} = 0, \\ 
 \int_{{\rho _v}}^{{\rho _l}} {\frac{1}{\rho }dp = \log (} {\rho _1}/{\rho _v}){\theta _v} + \log ({\rho _2}/{\rho _1}){\theta _m} + \log ({\rho _l}/{\rho _2}){\theta _l} = 0. \\ 
 \end{array}
 \end{equation}
 where $ {\rho _v} $
  and $ {\rho _l} $ are vapor and liquid coexistence densities, respectively. It is known that the equilibrium coexistence densities are completely determined by the mechanical stability condition for flat interfaces. However, for circular interfaces (e.g., droplets, bubbles), the Laplace's law also affects the coexistence densities. Due to the relatively large surface tension in the pseudopotential model (compared with the case in nature), the coexistence densities, especially the vapor phase density $ {\rho _v} $, usually change with the radius of curvature, and this density deviation is more significant for the large density ratio problems. For example, the vapor-density deviation can be as large as 60 $ \% $ for a system with $ {\rho _l}/{\rho _v}{\rm{ = }}100 $ \cite{li2014thermodynamic}. Li \textit{et al.} proposed that the density deviation can be much reduced by setting the vapor-phase sound speed $ \sqrt {{\theta _v}} $  to be the same the magnitude as $c_s$ \cite{li2014thermodynamic}. In addition, the interface thickness can be widened (sharpened) by decreasing (increasing) $\left| {{\theta _m}} \right|$. In this work, we consider the large density ratio problem with  $ {\rho _v} = 0.001 $ and  $ {\rho _l} = 1 $. The parameters  $ {\theta _v} $,   $ {\theta _l} $, and   $ {\theta _m} $ are given as
\begin{equation}\label{e21}
{\theta _v} = c_s^2/2,~~{\theta _l} = c_s^2,~~{\theta _m} =  - c_s^2/40.
\end{equation}
According to Eq. (\ref{e20}), the variables  $ {\rho _1} $ and  $ {\rho _2} $ are given as $ {\rho _1} = {\rm{0}}{\rm{.001325}} $  and  $ {\rho _2} = {\rm{0}}{\rm{.9758}} $. The parameter $ \sigma $  in Eq. (\ref{e17}) is set to 0.1 to achieve thermodynamic consistency.
\begin{figure}
	\center {
	\includegraphics[width=0.4\textwidth]{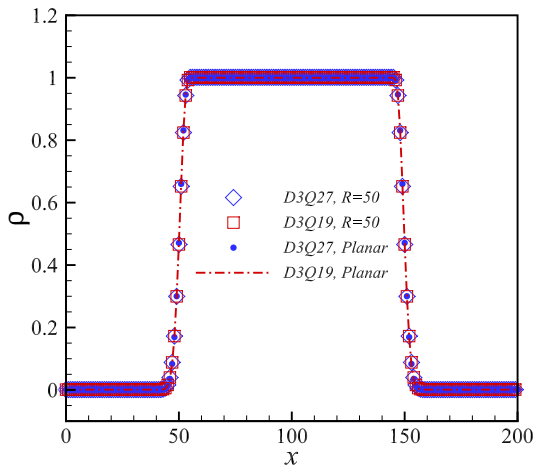}
}
	\caption{\label{FIG1} Comparison of the density profiles ( $ {\rho _v} = 0.001 $ and $ {\rho _l} = 1.0 $ ) at the Planar interfaces and circular interfaces ($ {R_0}=50 $) by the non-orthogonal D3Q27 and D3Q19 MRT-LBMs. For the planar interfaces, the two methods reproduce very accurate coexistence densities. For the circular interfaces, the small discrepancies in the vapor phase are within 6 $ \% $. The interface widths 
		obtained by Eq. (\ref{e23}) are  $ 5 \le W \le 6 $.}
\end{figure}

In the simulation, the square-root-form pseudopotential is used and the interaction strength parameter is fixed as $ G =  - 1 $. The density profiles along two planar interfaces in the $ x $ direction by the D3Q27 and D3Q19 non-orthogonal MRT-LBMs are shown in Fig. \ref{FIG1}. It is seen that the numerical coexistence densities are in very good agreement with the equilibrium vapor and liquid densities ( $ {\rho _v} = 0.001 $ and $ {\rho _l} = 1.0 $ ). A spherical droplet of radius ${R_0} = 50$ (a representative radius in the following applications) is then initially located at the center of a $ 200 \times 200 \times 200 $  cubic box to verify the thermodynamic consistency. The steady density profiles along the center line ($ y = 100,{\rm{ }}z = 100 $) are also shown in Fig. \ref{FIG1} for comparison. The liquid density at $ {R_0}=50 $ is basically the same as the value at the planar interface, while the small discrepancies in the vapor phase are within 6 $ \% $. Generally, the density ratio in our simulation is larger than 940. In addition, the interface width, $ W $, can be measured by fitting the following curve to the density profile,
\begin{equation}\label{e23}
\rho (x) = \frac{{({\rho _l} + {\rho _v})}}{2}{\rm{ + }}\frac{{({\rho _l} - {\rho _v})}}{2}\tanh \left[ {\frac{{2(x - 50)}}{W}} \right],~~x <  = 100.
\end{equation}
The above equation can be rewritten as  $W = ({\rho _l} - {\rho _v})/(\partial \rho /\partial x){|_{x = 50}}$ \cite{fei2018mesoscopic}. Using the numerical differentiation, the interface widths for density profiles in Fig. \ref{FIG1} are obtained as, $ 5 \le W \le 6 $.

 \subsection{Evolution of spurious velocities}
 Spurious currents are usually viewed as an important cause of instability in the pseudopotential model. Here the average spurious velocity magnitude in the gas phase
 ${{{{\bar u}}}_v}$ \cite{lycett2014multiphase} is considered  to compare the numerical performances of the proposed non-orthogonal MRT-LBMs and the classical SRT-LBM. The SRT-LBM is obtained by setting all the relaxation parameters equal to one another. The  spurious velocity is measured based on the static droplet case in sec. \ref{sec.3a}
 at different viscosities. It may be noted that the dynamic viscosity ratio is $ {\mu _l}/{\mu _v} = ({\rho _l} /{\rho _v} ) \approx 1000 $  in this simulation due to the unity kinematic viscosity, while different dynamic viscosity ratios can be used in the following applications.
 
\begin{figure}
	\center {
		\includegraphics[width=0.4\textwidth]{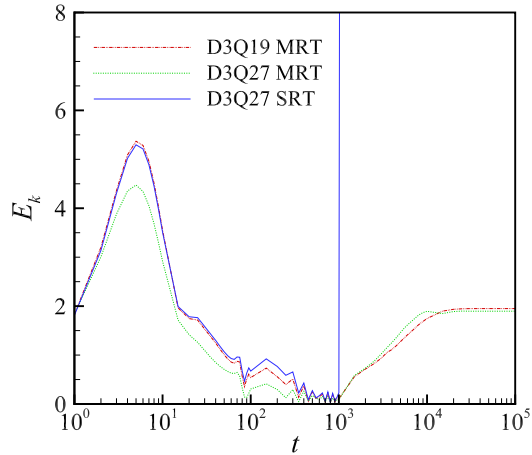}
	}
	\caption{\label{kinetic_energy}
		Evolution of spurious kinetic energy ${E_k}$ for a static droplet with $\nu  = 0.02$.
	 }
\end{figure}

 From Table \ref{TAB1}, it is seen that the present non-orthogonal MRT-LBM models help to reduce the spurious currents compared with SRT-LBM, and the D3Q27 model outperforms the D3Q19 model at low viscosities. In addition, we also provide the evolution of spurious kinetic energy for the case $\nu  = 0.02$ in Fig.~\ref{kinetic_energy}. Here the spurious kinetic energy is calculated based on the global integral of the spurious velocities.
 The SRT-LBM case  diverges after
 1000 steps, while the present models allow us to achieve convergent results. Clearly, the proposed non-orthogonal MRT-LBM has superior numerical stability over the SRT-LBM.
 \begin{table*}
 	\center {
 	\renewcommand\arraystretch{1.2}
 		\caption{\label{TAB1}%
 		Average spurious velocity magnitude in the gas phase ${{{{\bar u}}}_v}$  produced by different methods.
 		}
 		\begin{tabular}{ccccc}
 			\toprule
 			 methods &$\nu  = 0.075$&$\nu  = 0.05$&$\nu  = 0.02$&$\nu  = 0.01$\\
 			\hline
 			 D3Q27 MRT  & 0.00116& 0.00293 & 0.0172&0.0245\\			
 			 D3Q19 MRT   & 0.00085 & 0.00333 & 0.0180& 0.0295\\
 			 D3Q27 SRT   & 0.0150 & 0.06542 & NaN & NaN\\
 			 \toprule
 	\end{tabular}
 }
 \end{table*}
 \subsection{Realization of tunable surface tension}
\begin{figure}
	\center {
	\includegraphics[width=0.32\textwidth]{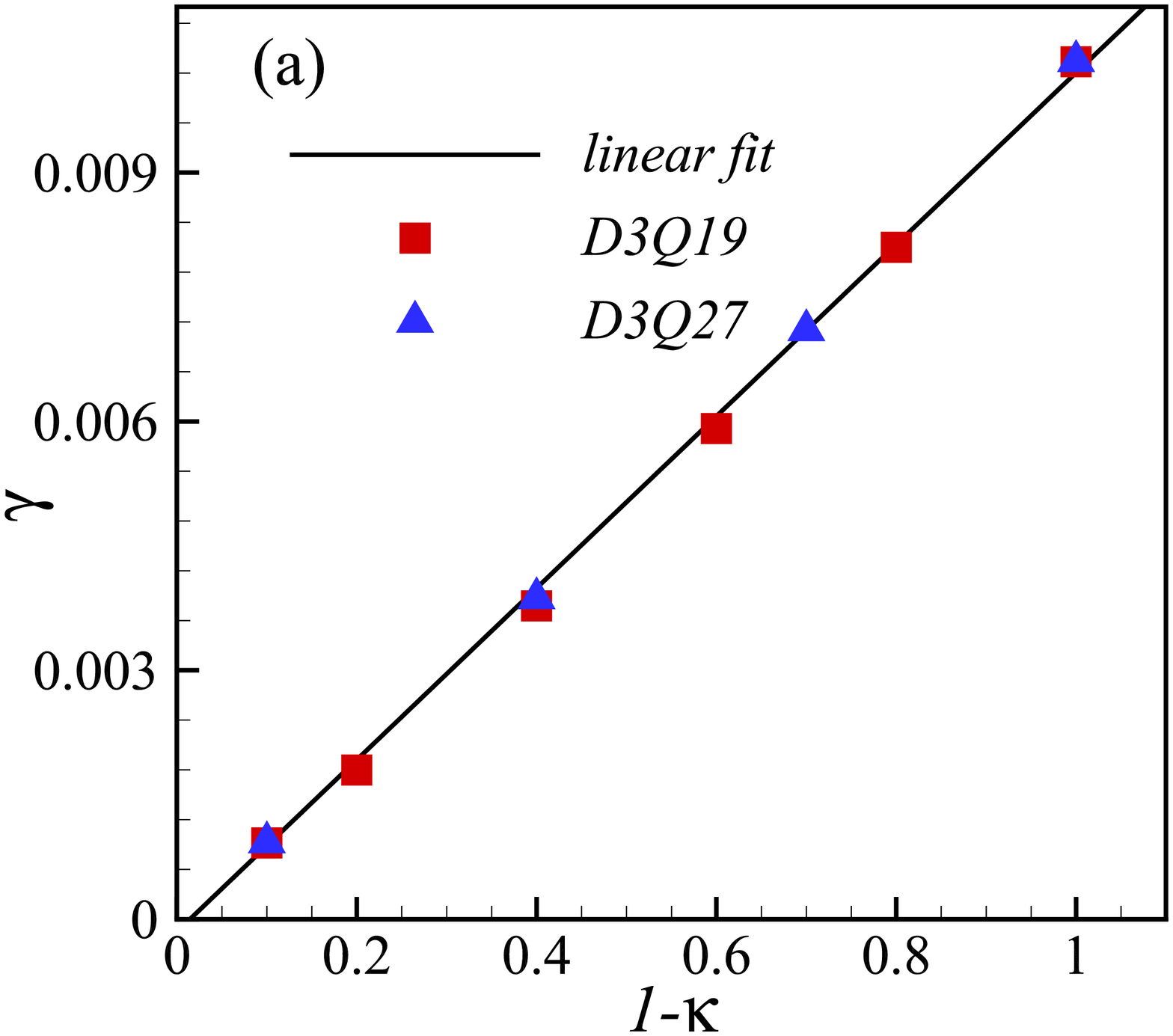}
	\includegraphics[width=0.32\textwidth]{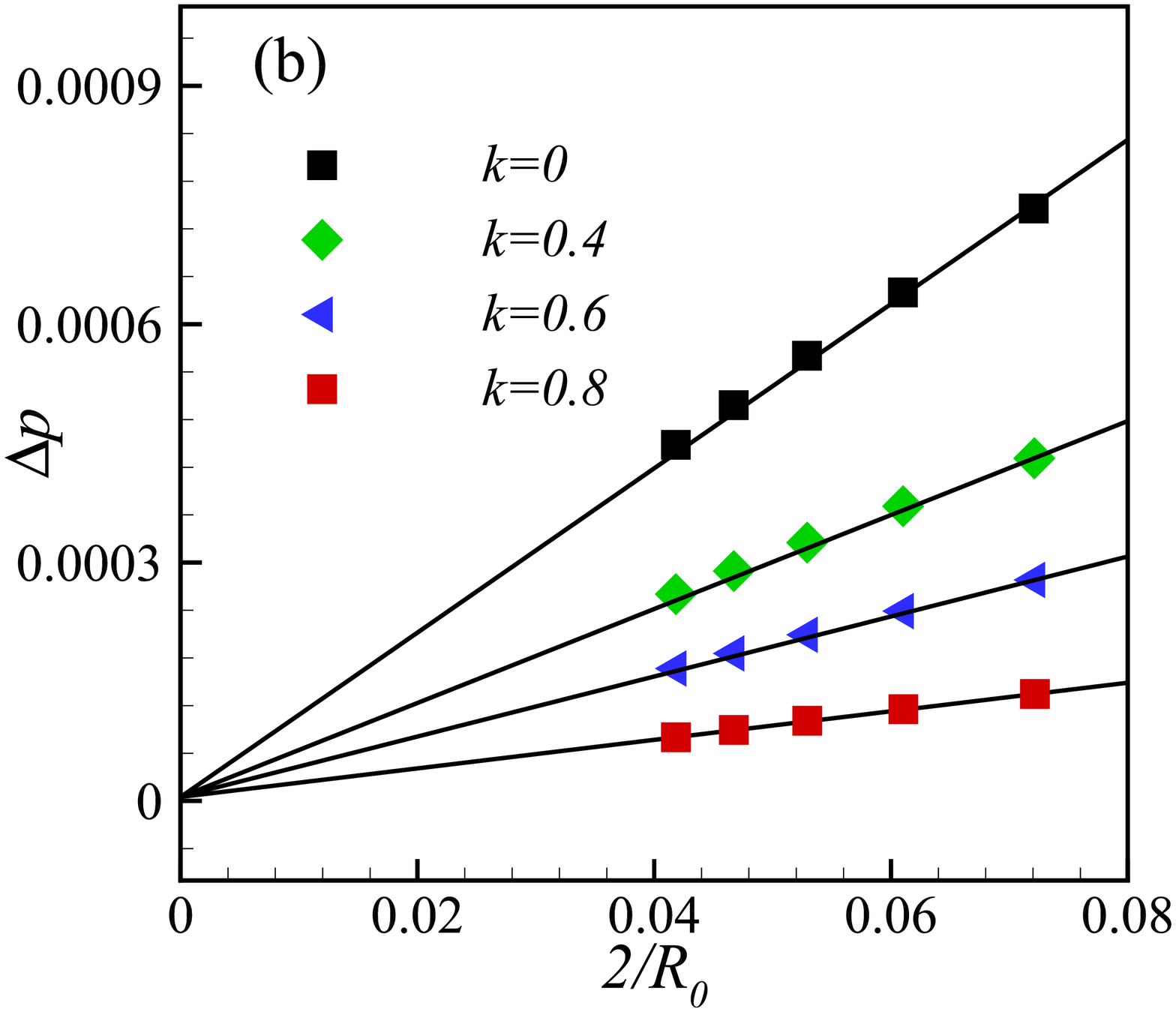}
	\includegraphics[width=0.32\textwidth]{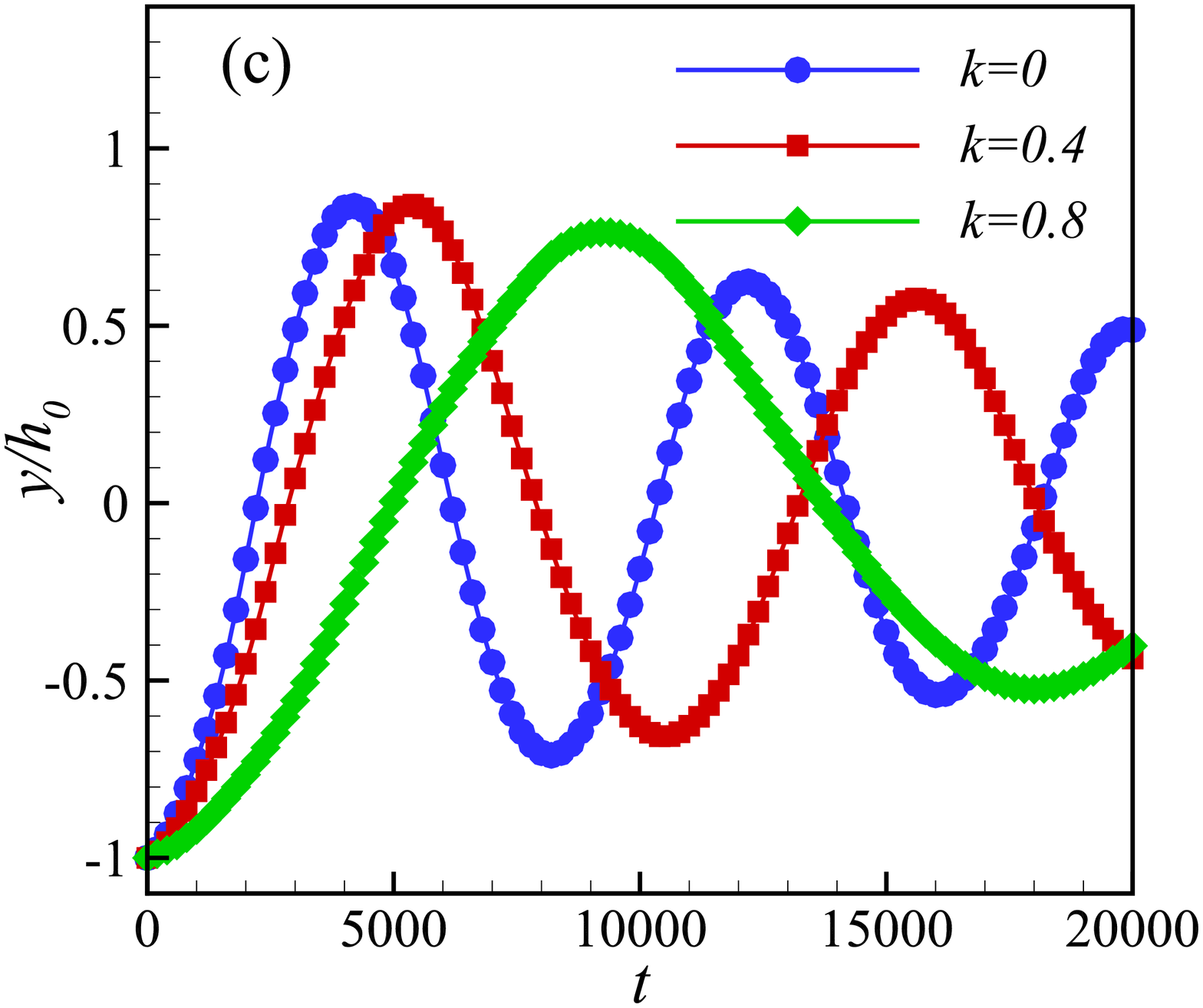}	
}

	\caption{\label{FIG2} Verification of the surface tension adjustment method in Eq. (\ref{e18}). (a) and (b) are based on the static droplet case: (a) surface tensions achieved at different $ k $; (b) numerical pressure difference as a function of $2/{R_0}$
		at different $ k $ by D3Q19 non-orthogonal MRT model.  (c) Dynamic decay of capillary waves: for  $k = [0,0.4,0.8]$, the analytical oscillation periods based on the surface tension by the static cases are $T = [7959, 10504, 19056]$, while the measured periods are ${T^*} = [8025,10400,18500]$, respectively.		
	}
\end{figure}

The adjustment of the surface tension is first verified by simulating droplets with a series of radii, $ {R_0} = [28,33,38,43,48] $ , in a $ 180 \times 180 \times 180 $  cubic box. According to the Laplace's law, the pressure difference across a spherical interface is related to the droplet radius $ {R_0} $ and the surface tension $ \gamma $  via $ \Delta p = {p_{{\mathop{\rm int}} }} - {p_{out}} = 2\gamma /{R_0} $.
The obtained surface tensions at different $ k $ are shown in Fig. \ref{FIG2}a. As is shown, there is a good linear scaling between the surface tension and (1-$\kappa $), which confirms the theoretical analysis that surface tension can be reduced linearly with increasing
parameter $ k $ in Eq. (\ref{e18}). In addition, the numerical pressure differences at $\kappa $=0, 0.4, 0.6, and 0.8 by D3Q19 non-orthogonal MRT model are given in Fig. \ref{FIG2}b. It can be seen that the numerical results agree well with the linear fit denoted by the solid lines. In the following simulations, we only adopt the D3Q19 model due to its smaller computational load.

In addition to the static case,  we consider the decay of capillary waves between two fluids with equal viscosities ($\nu  = 0.01$), which is a classical test for the accuracy of numerical models for surface-tension-driven interfacial dynamics \cite{shan1994simulation,li2013achieving}. The computational domain is in a cuboid of length $ L $, height $ H $ and depth $ D $. For convenience, we use $ D=5 $ and impose periodic condition in the $ z $ direction. As suggested in \cite{shan1994simulation,li2013achieving}, the aspect ratio $ H/L $ should be large enough and is chosen as 5 with $ L=160 $. The periodic and nonslip boundary conditions are adopted in the $ x $ and $ y $ directions, respectively. Initially, an interfacial disturbance is given in the middle of the cuboid of the form $y(x) = {h_0}\cos (kx)$, where $k = 2\pi /L$ is the wave number and $ {h_0}=20 $ is the wave amplitude. For the given surface tension, the dispersion relation of capillary wave is \cite{shan1994simulation},  ${\Theta ^2} = \gamma {k^3}/({\rho _l} + {\rho _v})$. 
Figure \ref{FIG2}c shows the evolution of the interface at $ x=L/2 $ for three cases by the present non-orthogonal MRT-LBM. We can clearly see that the dynamic decay of capillary waves can be well captured using the proposed method, and the  oscillating period increases with the decrease of surface tension (increase of $ k $). To be quantitative, we compare the measured oscillating period ${T^*}$ with the theoretical value $T = 2\pi /\Theta $. Generally, the present results are in very good agreement with the analytical results, with relative errors of 
$ 0.8\% $, $ 1.0\% $ and $ 2.9\% $ for $ k=0 $, $ k=0.4 $ and $ k=0.8 $, respectively.

\subsection{Validation of spatial accuracy}
\begin{figure}
	\center {
		\includegraphics[width=0.4\textwidth]{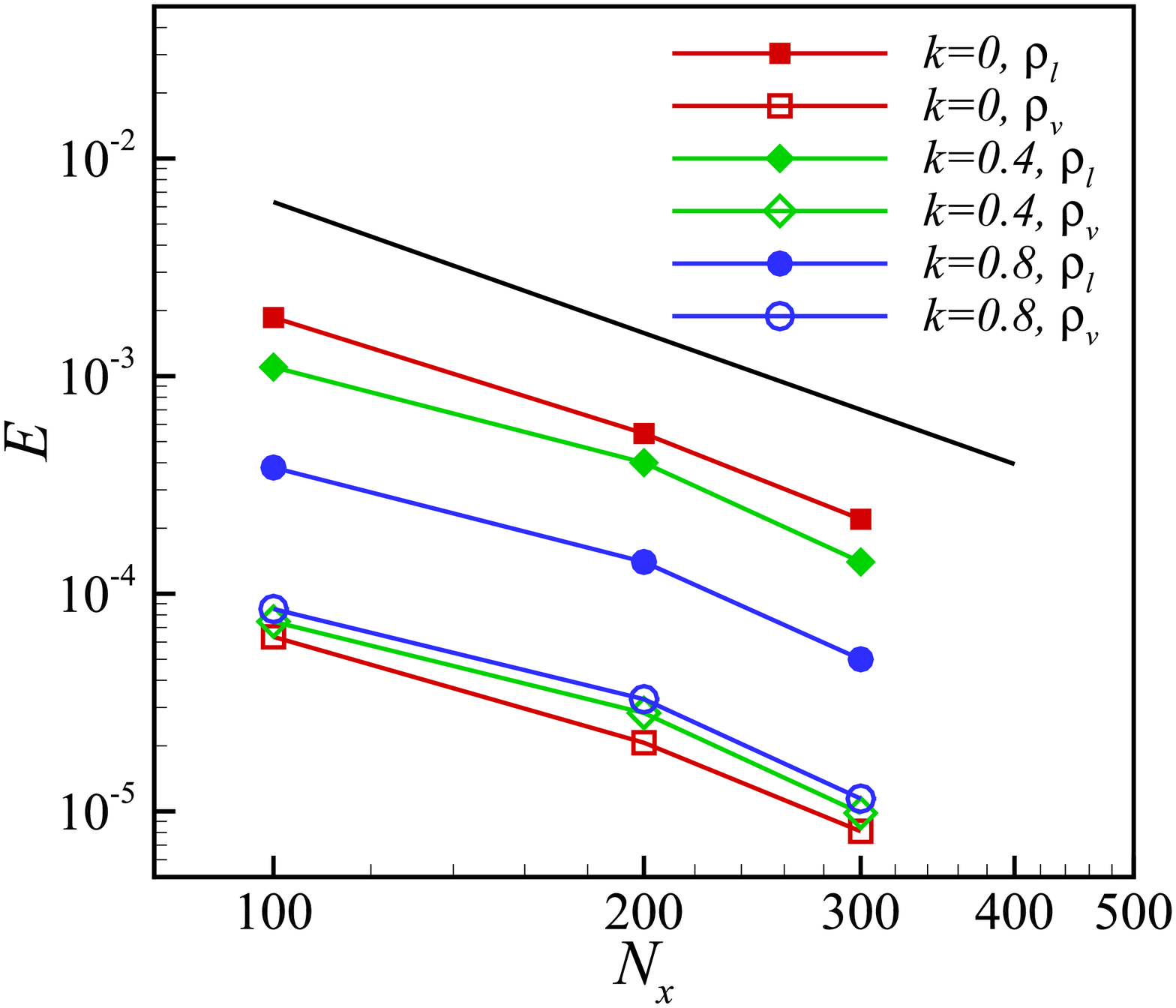}
	}
	\caption{\label{convergent}
			Changes in the relative error $ E $ of steady densities with the mesh size ${N_x}$. The black solid line represents exact second-order accuracy. Three cases with different surface tensions ($ k $) are considered.
		}
\end{figure}

To test the spatial accuracy of the proposed non-orthogonal MRT-LBM for multiphase flow, we conduct simulations of a static droplet 
with different mesh sizes, ${N_x} \times {N_y} \times {N_z} = 100 \times 100 \times 100$, $ 200 \times 200 \times 200 $, $ 300 \times 300 \times 300 $, and $ 400 \times 400 \times 400 $. The droplet radius is $R = {N_x}/4$ and the periodic boundary conditions are imposed in all three directions. As suggested in \cite{xu2015three}, we use the finest mesh as the standard case and calculate the relative error for the results on other meshes by $E({N_x}) = \left| {\rho ({N_x}) - \rho (400)} \right|$, where ${\rho ({N_x})}$
represents the convergent value of liquid/gas density on the mesh ${N_x} \times {N_y} \times {N_z}$. The changes in the relative error with the mesh size for different values of $ k $ are shown in Fig.~\ref{convergent}, where the top black line stands for the exact second-order accuracy. It is  demonstrated that the present model has approximately second-order accuracy in space.

\subsection{Implementation of wettability conditions}
Next, we consider the implementation of the wettability. For the pseudopotential model, various schemes to implement the contact angle between the fluid and solid phases have been proposed in the literature \cite{huang2007proposed,li2014contact}. In this work, we adopt a modified pseudopotential-based contact angle scheme \cite{li2014contact}, in which a fluid-solid interaction is defined as,
\begin{equation}
{{\bf{F}}_{ads}}({\bf{x}}) =-{G_{ads}}\psi ({\bf{x}})\sum\limits_i {w({{\left| {{{\bf{e}}_i}} \right|}^2})} \psi ({\bf{x}})s({\bf{x}} + {{\bf{e}}_i}\Delta t){{\bf{e}}_i}
\end{equation}
where $ {G_{ads}} $  is the fluid-solid interaction strength to adjust the contact angle, and $ s({\bf{x}}) $  is an indicator function, which is equal to 1 or 0 for a solid or a fluid phase, respectively. For such a treatment, $ {G_{abs}} < 0 $ , $ {G_{abs}} = 0 $ , and $ {G_{abs}} > 0 $  recover the hydrophilic, neutral, and hydrophobic walls, respectively.  However, there is still no analytical relation between the specified contact angle and the value of $ {G_{ads}} $ . Usually,  $ {G_{ads}} $ is set to match the prescribed one. In the present work, the intrinsic contact angles are implemented without considering contact angle hysteresis. For cases where the three-phase contact line motion is a dominant factor \cite{Eddi2013,Aashutosh2018}, alternative contact angle schemes, such as the geometric formulation \cite{Dinghang2007,Wanglei2013}, should be adopted to include the contact angle hysteresis.

As a benchmark case, we choose  $ {G_{abs}}=0.23 $, the measured contact angle is around  $ {157^ \circ } $, which is a representative value for super-hydrophobic surfaces. To verify the implementation, a droplet impact on a solid wall is simulated. In this problem, the droplet spreads firstly to reach a maximal spreading diameter, then retracts to reduce its interfacial energy, and finally rebounds from the solid surface due to the relatively small energy loss by
dissipation and friction.  According to the universal scaling summarized by Richard  \textit{et al.}  \cite{richard2002surface}, the contact time $ {t_c} $ , a time period from when the droplet first touches the surface to that when it bounces off the surface, is proportional to the inertia-capillarity time,
\begin{equation}\label{e24}
\tau  = \sqrt {{\rho _l}{{R_0}^3}/\gamma }, 
\end{equation}
where the scale factor $ {t^*} = {t_c}/\tau  \approx 2.2 \pm 0.3 $  is independent of the impact velocity  $ U $ and holds in a range of the Weber number, $ We = {\rho _L}{R_0}{U^2}/\gamma $ . 

\begin{figure}
	\center {
	\includegraphics[width=0.45\textwidth]{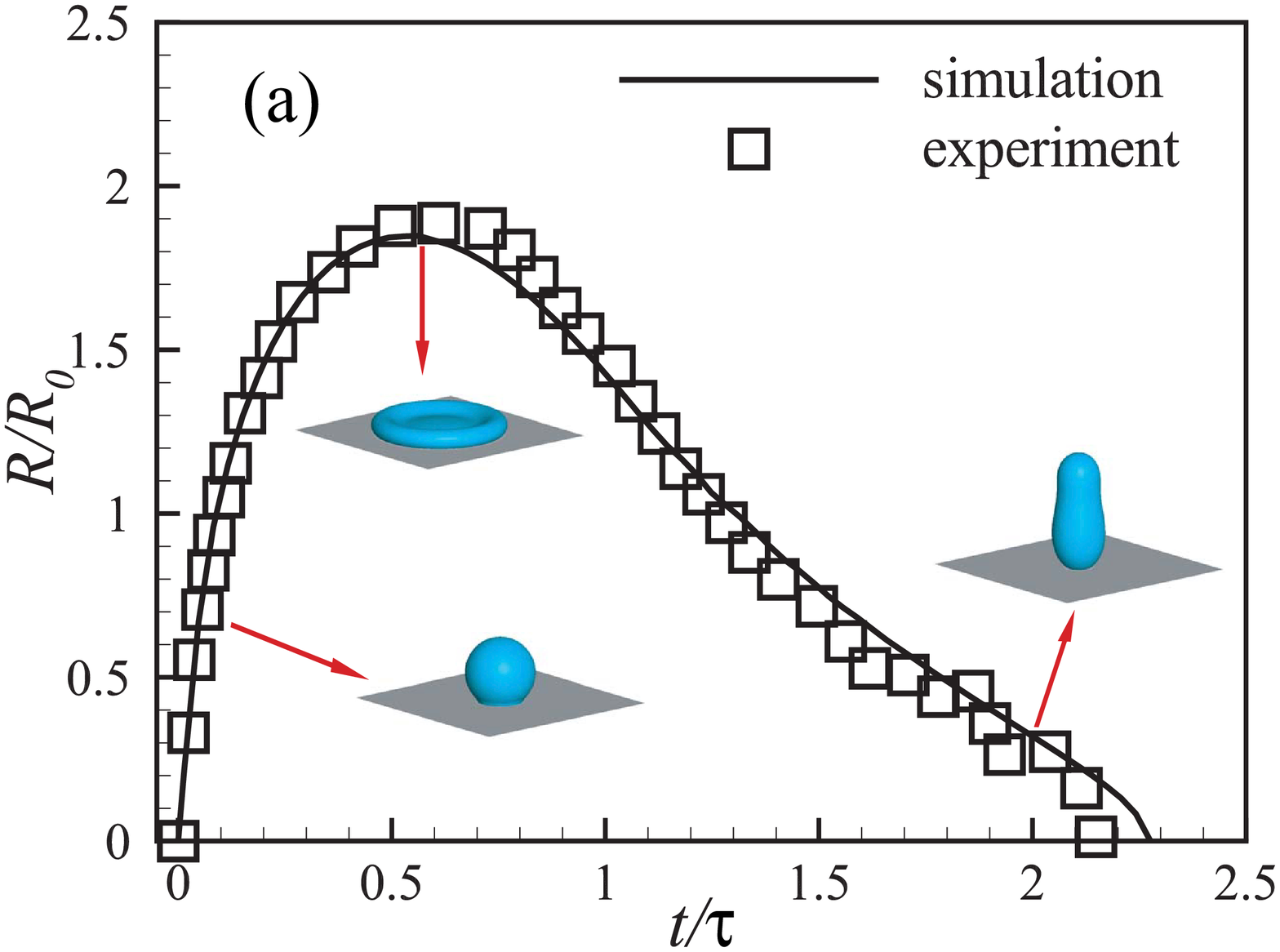}
	\includegraphics[width=0.45\textwidth]{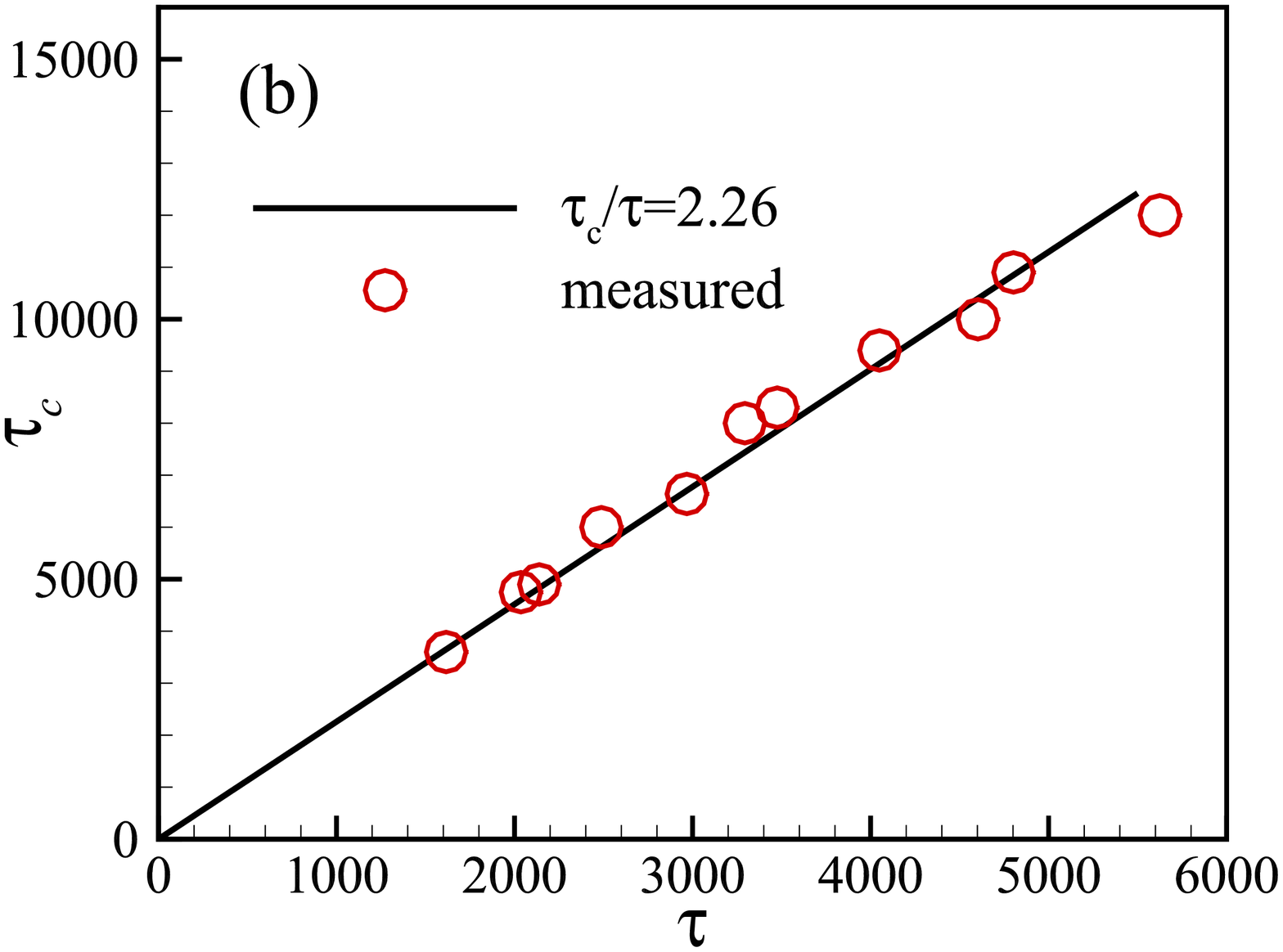}
	}
	\caption{\label{FIG4} (a) Time evolution of the dimensionless film  radius for a water droplet impacting on the superhydrophobic surface at	$We = 27$. Line: present simulation; symbols: experiment in \cite{bird2013reducing}. Inset: typical snapshots during the impacting process. (b) Contact time ${\tau _c}$ as a function of the inertia-capillarity  time $\tau $ for a droplet impacting on the superhydrophobic surface.
	}
\end{figure}

We consider a series of cases with the droplet radius $ 30 \le {R_0} \le 50 $ and surface tension  $ 0.00395 \le \gamma  \le 0.01034 $ ($ 0 \le \kappa  \le 0.6 $). The resulting range of the inertia-capillarity time is $1616 \le \tau  \le 5625$ . The simulations are run in a domain of dimensions around  6$ {R_0} $. In Fig.~\ref{FIG4}a, we show the time evolution of the spreading film radius $R$ for a representative case with $ W{\rm{e}} =27 $, where we can see the present simulation is in good agreement with the experimental result reported in 
\cite{bird2013reducing}. The contact time as a function of the inertia-capillarity time is shown in Fig. \ref{FIG4}b. Within the considered range of parameters, a very good scaling  $ {\tau _c}/\tau {\rm{ = }}2.26 $ is achieved, which validates our implementation of wettability conditions, as well as the adjustment of surface tension.

\section{Numerical applications in realistic multiphase flows}\label{sec.4}

\subsection{Fuel droplet impact on a solid surface}\label{sec.3b}
Fuel droplet impact on a solid surface occurs in fuel spray in engines, spray cooling, and inking jet printing \cite{moreira2010advances}. As discussed in the literature, the impact outcome depends on the properties of both the fuel droplet and the surface \cite{moreira2010advances}. 
Here we consider the ethanol droplet and diesel fuel droplet impact on hydrophilic surface. The simulation configuration is that a droplet with radius $ {R_0} $ and velocity $U$  impact on the solid wall vertically (in the $ z $ direction). The simulations results by the present non-orthogonal MRT-LBM are compared with the experimental data provided in Ref.
\cite{moita2007experimental} and a recent smoothed particle hydrodynamics (SPH) simulation \cite{yang2017simulation}. In the present work, the effect of temperature, such as the phase change process during drop impact on heated surfaces\cite{Yang2018SPH}, is not considered. For methods of incorporating thermal effects into multiphase LBM, the interested readers are kindly directed to Ref. \cite{li2016lattice}.

The physical properties of the ethanol and diesel have been provided in Ref. \cite{moita2007experimental}. Due to the large density ratio for both the two fuel droplets compared with the ambient gas. The density ratio is fixed to be the same with the previous setting. The experimental dynamic viscosity ratio $ {\mu _l}/{\mu _v} $ can be achieved by tuning the vapor-liquid kinematic viscosity ratio $ {\nu _l}/\nu {_v} $ through a variable relaxation time, i.e.,
\begin{equation}
\nu  = {\nu _v} + ({\nu _l} - {\nu _v})\frac{{\rho  - {\rho _v}}}{{{\rho _l} - {\rho _v}}}.
\end{equation}
Remarkably, the impact Reynolds number ${\rm{Re}} = {\rho _l}{R_0}U/{\mu _l}$
and Weber number $We = {\rho _l}{U^2}{R_0}/\gamma $ in the present method can be tuned independently via the adjustments of the surface tension and the viscosity to match the experimental conditions. It may be noted that the Reynolds and Weber numbers are defined based on the droplet radius
 $ {R_0} $  throughout this paper, while the droplet diameter $ D=2{R_0} $ has also been commonly adopted in the literature. The fluid-solid interaction strength is set to  ${G_{ads}} =  - 0.1$ to match the wettability condition \cite{yang2017simulation}.

\begin{figure}
		\center {
		\includegraphics[width=0.48\textwidth]{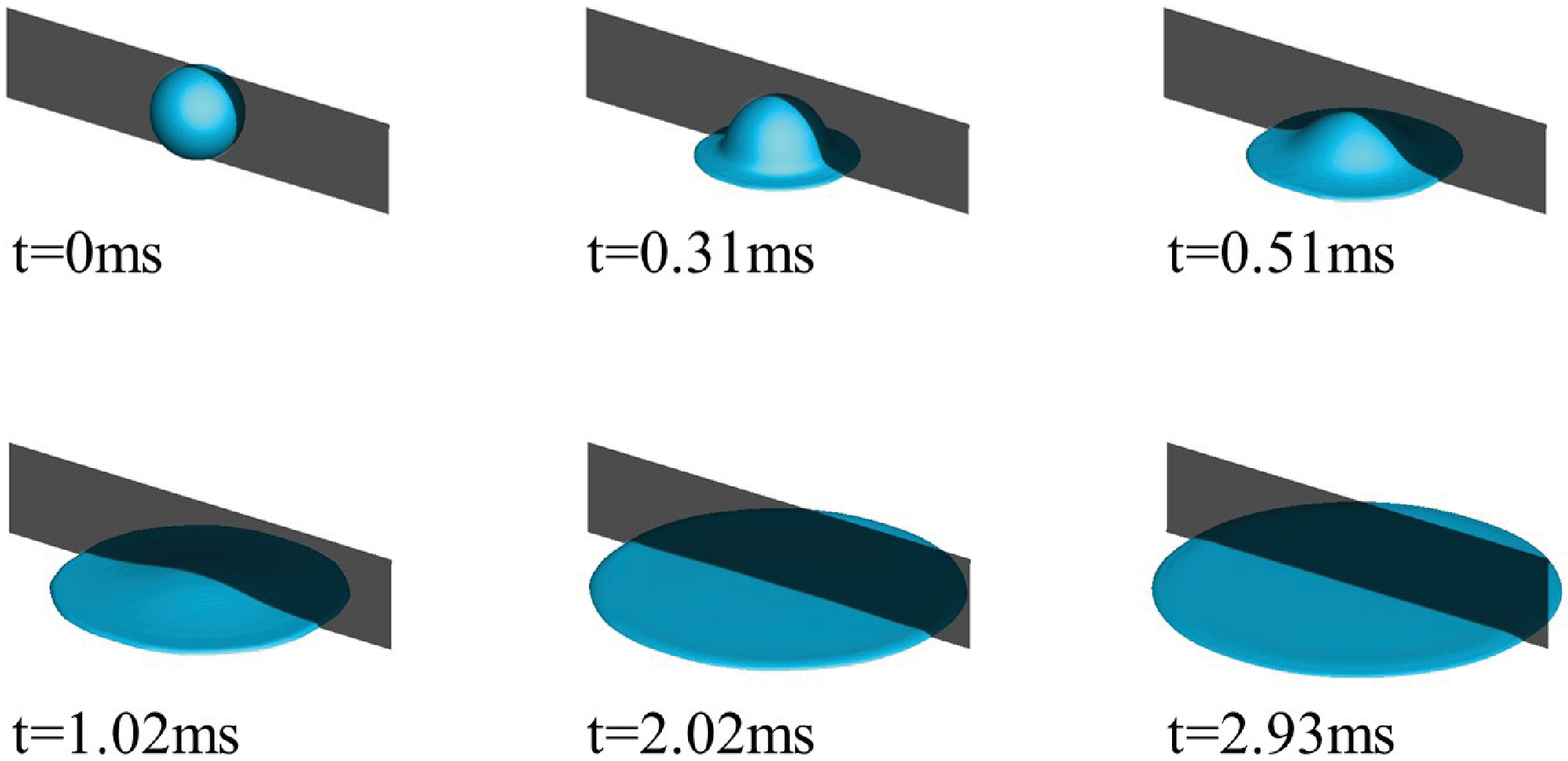}\\
		(a)\\
		\includegraphics[width=0.35\textwidth]{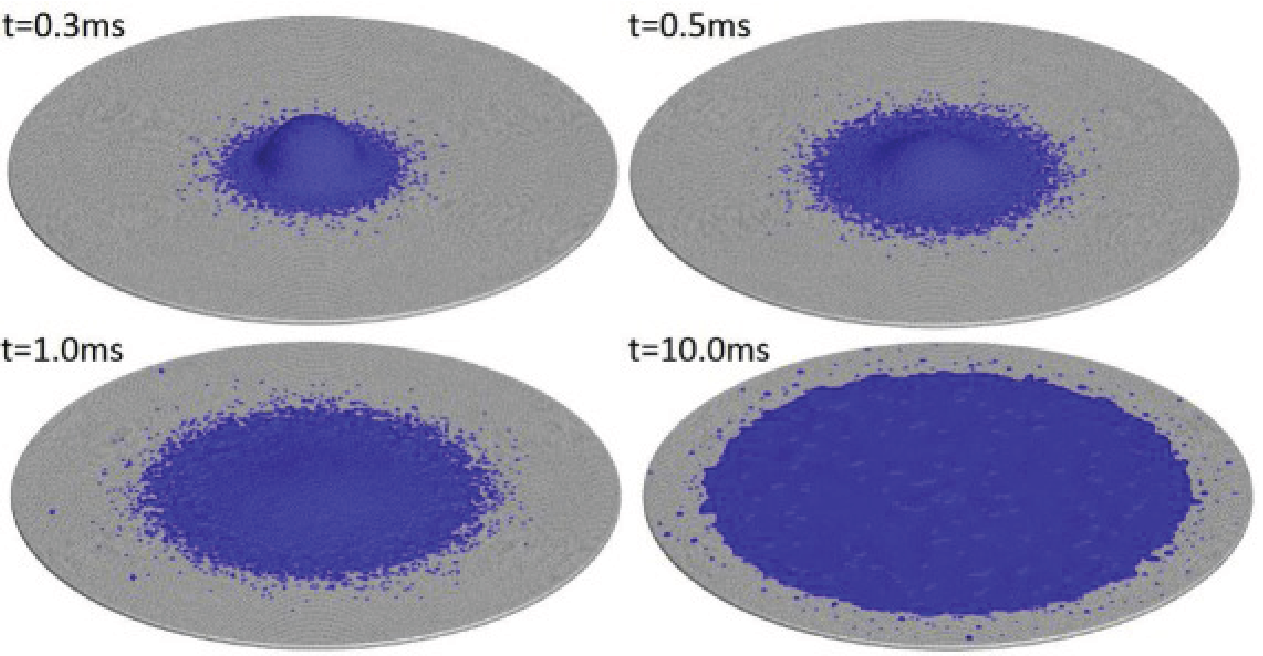}\\
		(b)
	}
	\caption{\label{FIG5} Snapshots of an ethanol drop impinging on a drying surface (${R_E} = 1.2mm$, ${U_E} = 3.1m/s$): (a) the present non-orthogonal MRT-LBM simulation; (b)  SPH simulation in \cite{yang2017simulation}.
	}
\end{figure}

The first simulation is an impact case by the ethanol drop. In the experiment, the droplet radius and impact velocity are  ${R_E} = 1.2mm$ and ${U_E} = 3.1m/s$. According to the physical properties \cite{yang2017simulation},
the corresponding Reynolds and Weber numbers are around  ${\mathop{\rm Re}\nolimits}  = 2500$ and $We = 410$, respectively. In the simulation, we choose ${R_0} = 70$, $U = 0.125$ and ${\nu _l} = 0.0035$. The simulation is run in a domain around $9R \times 9R \times 3.5R$, where the periodic boundary conditions are used in the $ x $ and $ y$ directions and the non-slip boundary conditions on the top and bottom walls. Figure \ref{FIG5} shows the predicted evolution of the droplet impacting process. Specifically, the liquid droplet spreads out onto the solid wall and gradually forms a thin liquid film without splashing, which is consistent with the experimental observation \cite{moita2007experimental}. For comparison, the SPH simulation result in \cite{yang2017simulation} is also shown in Figure \ref{FIG5}. As expected,  the present film is smoother than the SPH simulation.
It should be noted that the rough boundaries in the film by SPH are not the secondary droplets due to splashing, as mentioned in \cite{yang2017simulation}. 

\begin{figure}
	\center {
	\includegraphics[width=0.4\textwidth]{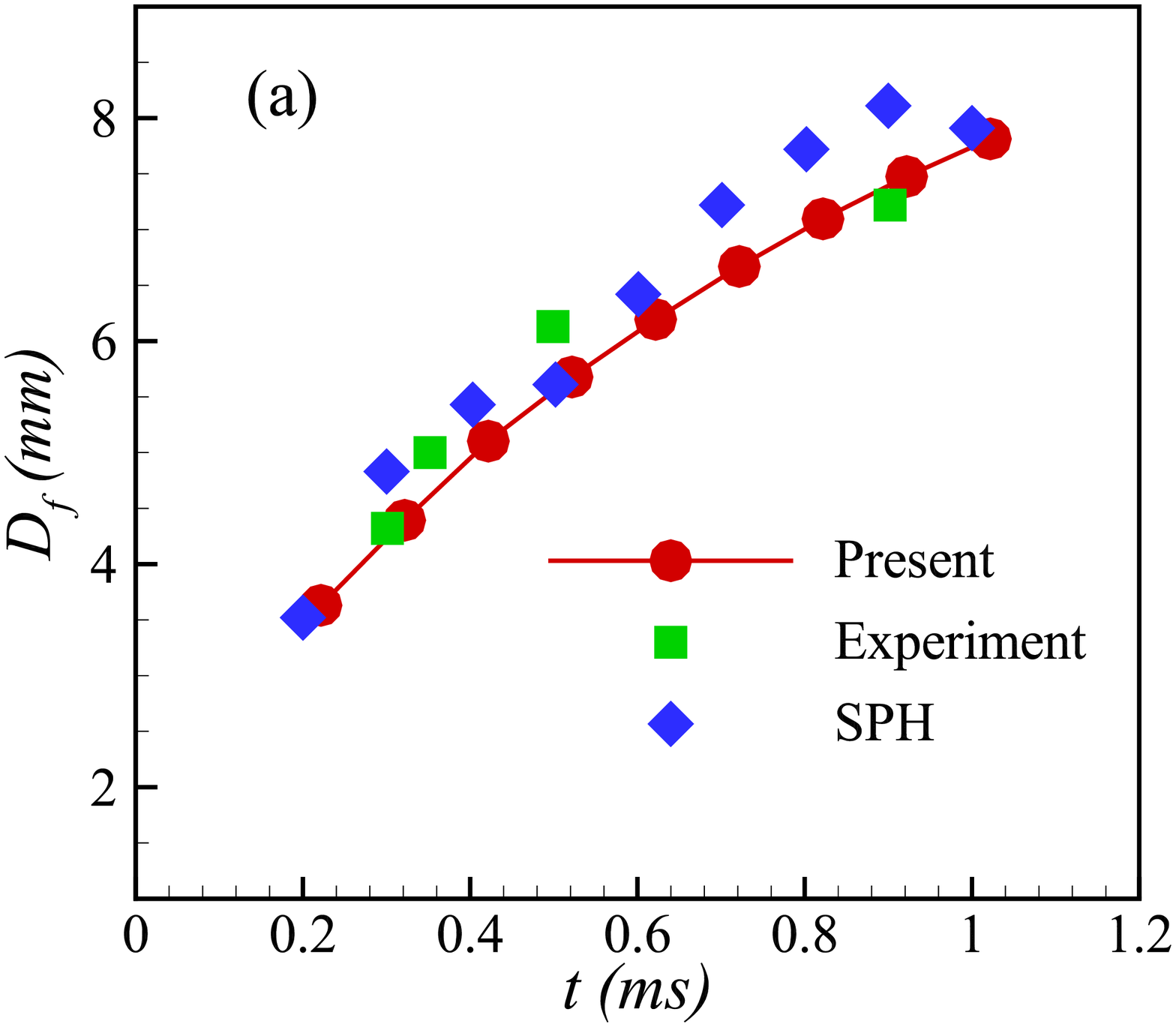}
	\includegraphics[width=0.4\textwidth]{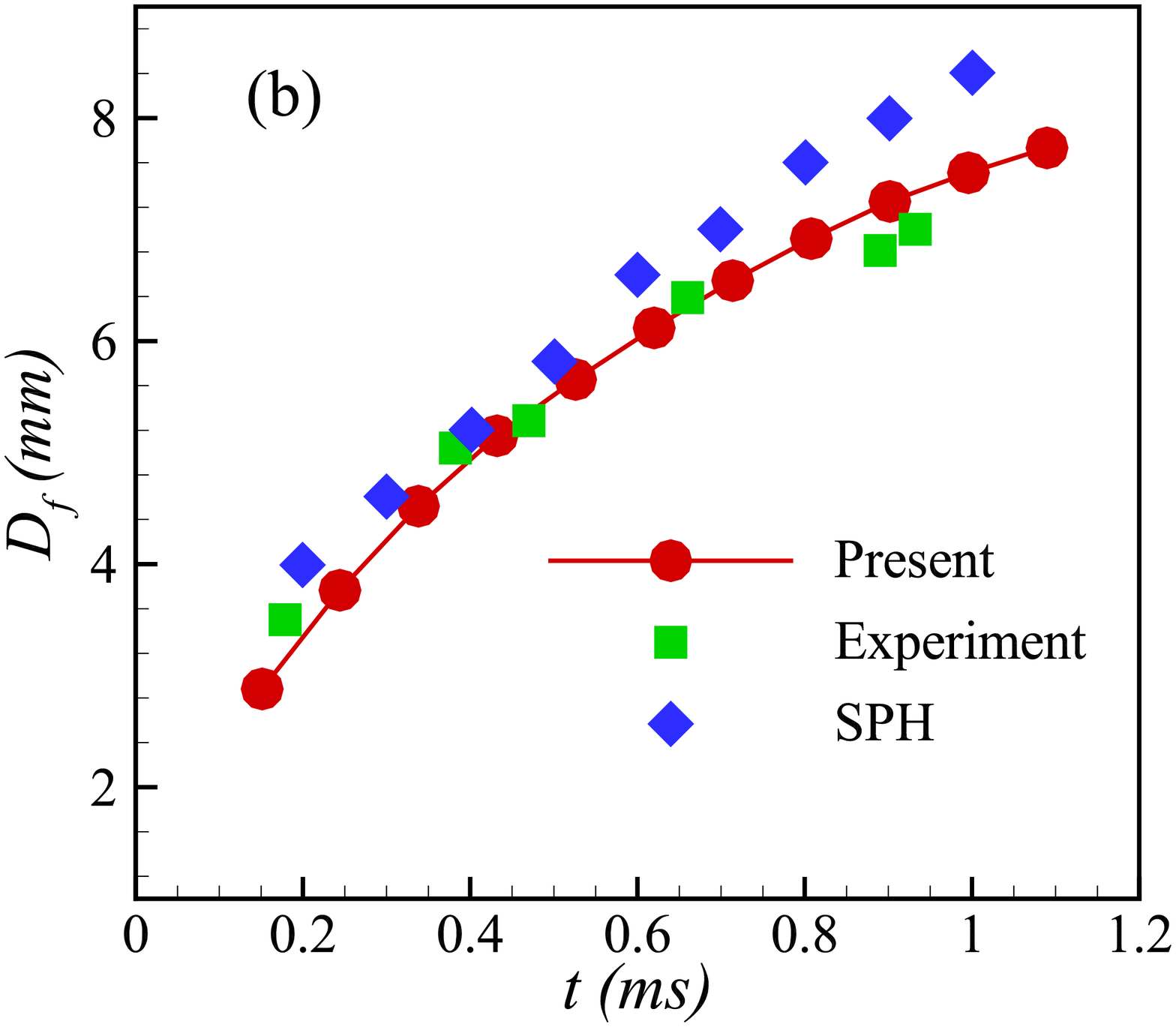}
	}
	\caption{\label{FIG6} The spreading diameters as a function of time by the present simulation, experimental measurement \cite{moita2007experimental} and SPH simulation \cite{yang2017simulation}: (a) ethanol drop impact (${R_E} = 1.2mm$ and ${U_E} = 3.1$); (b) diesel drop impact (${R_E} = 1.3mm$ and ${U_E} = 3.1$).
	}
\end{figure}

The diameter of the spreading film ${D_f}$ is then measured. To compare with the experimental data, we can convert the lattice time 
and spreading film diameter to the experimental units using the dimensionless 
time and length scales, ${t^*} = tU/{R_0}
$ and ${\delta ^*} = {D_f}/{R_0}$. The predicted spreading diameter, as a function of time, compared with the experimental measurement and SPH simulation, is shown in Fig. \ref{FIG6}. In addition, another impact case for diesel droplet (${R_E} = 1.3mm$ and ${U_E} = 3.1$) is also shown in Fig. \ref{FIG6}. Due to the smaller Reynolds and Weber numbers (${\mathop{\rm Re}\nolimits}  = 930$ and $We = 350$), ${R_0} = 50$ is used to simulate the diesel droplet impingement and a video for this case is provided in the supplementary movie 1. It can be seen in Fig. \ref{FIG6} that the present simulation results are generally in good agreement with the previous data \cite{moita2007experimental,yang2017simulation}. More specifically, the spreading diameters by the present method are smaller than the SPH results in the later stages
and seem to be more consistent with the experimental results.

In addition, we find that for the cases considered here with large Reynolds and Weber numbers, the SRT scheme always leads to divergence shortly after the droplet lands on the wall, which further confirms the improved numerical stability of the proposed non-orthogonal MRT scheme.

\subsection{Droplet impact on a super-hydrophobic wall with a cylindrical obstacle}\label{sec.3c}
Reducing the contact time for a droplet impact on a super-hydrophobic solid surface plays an important role in a broad range of realistic applications, such as self-cleaning, anti-icing and dropwise condensation \cite{liu2015symmetry,liu2014pancake,andrew2017variation}. Recently, several methods have been demonstrated to reduce the contact time \cite{liu2015symmetry,liu2014pancake,gauthier2015water}. In this paper, we consider the approach by installing a cylinder on a  super-hydrophobic solid surface, while the cylinder has the same wettability
with the solid surface. As analyzed by Liu \textit{et al.} \cite{liu2015symmetry}, when a droplet lands, more momentum is transferred into the azimuthal direction of the cylinder rather than the axial direction. As a result, the droplet remains extended in the azimuthal direction when it begins to retract along the axial direction. It is the dynamically asymmetric momentum and mass distribution that reduces the total contact time during the impact process.
\begin{figure}
	\center {
		\includegraphics[width=0.48\textwidth]{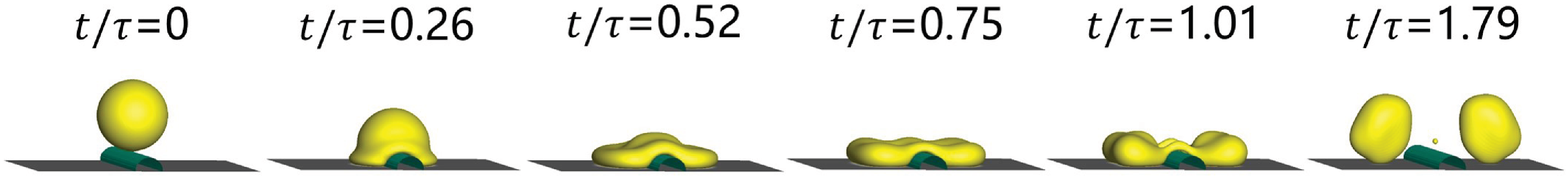}\\
		(a)\\
		\includegraphics[width=0.48\textwidth]{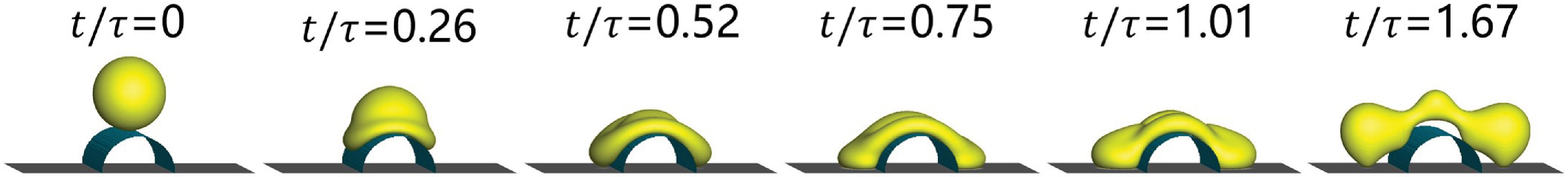}\\
		(b)\\
		\includegraphics[width=0.48\textwidth]{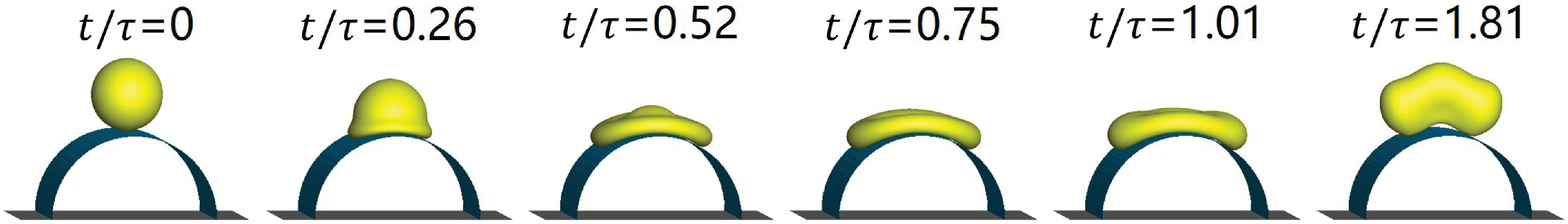}\\
		(c)
	}
	\caption{\label{FIG7} Snapshots for a droplet impact on a super-hydrophobic surface with a cylindrical obstacle by the present simulation: (a)	$R/{R_0} = 0.4$, rebounce at
		${t^*} \approx 1.79$; (b)	$R/{R_0} = 1.0$, rebounce at
		${t^*} \approx 1.67$; (c)	$R/{R_0} = 2.3$, rebounce at
		${t^*} \approx 1.81$.
	}
\end{figure}
\begin{figure}
	\center {
		\includegraphics[width=0.48\textwidth]{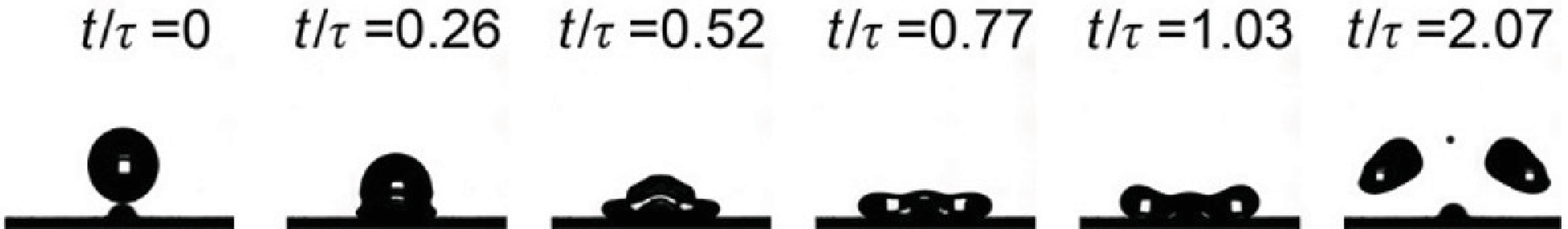}\\
		(a)\\
		\includegraphics[width=0.48\textwidth]{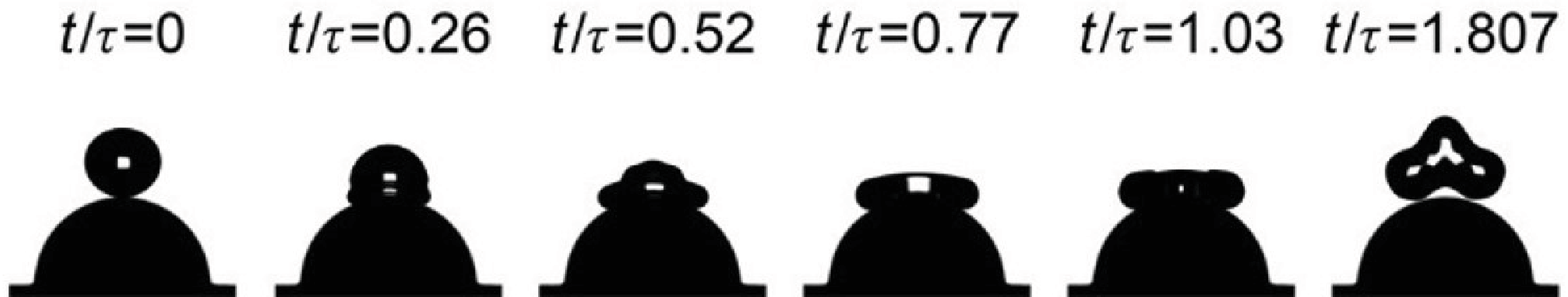}\\
		(b)
	}
	\caption{\label{FIG8} Experiment measurements for a droplet hitting  a cylinder \cite{liu2015symmetry,andrew2017variation}: (a) $R/{R_0} = 0.4$ and (b) $R/{R_0} = 2.3$.
	}
\end{figure}

In the simulation, the droplet radius is fixed at $ {R_0} = 50 $, and a series 
cases with the cylinder radii $ 10 \le R \le 120 $ are considered. The simulations are carried out in a box around $ 6{R_0} \times 6{R_0} \times 5{R_0} $, with periodic boundaries along $ x $ and $ y$ directions, and non-slip boundary conditions on the top and bottom, as well as the cylinder surface. The liquid viscosity is set to $ {\nu _l} = 0.0075 $ and $ Oh = ({\rho _l}{\nu _l})/\sqrt {{\rho _l}\gamma {R_0}}  < 0.015 $ for two Weber numbers, $ We = 10 $ and $ 20 $. The gravity is neglected since all the simulations are below the inertial capillary length scale. The fluid-solid interaction parameter ${G_{ads}} = 0.23$ is used to implement a static contact angle $ \theta  \approx {157^ \circ } $. A reference case without the cylinder is simulated first and the dimensionless contact time is $ t/\tau  \approx 2.35 $, which is in good agreement with the universal scaling \cite{richard2002surface}.

Figure \ref{FIG7} shows snapshots for three representative cases, $ R < {R_0} $, $ R={R_0} $, and $ R > {R_0} $ at $ We = 10 $. For the case $ R/{R_0} = 0.4 $, the droplet spreads onto the plat surface soon after the impact due to the small cylinder radius. Thus only a small part of the droplet is on the cylinder ridge and this small central part retracts quickly while the main part is still spreading. This results in a central pinch-off, splitting the droplet into two parts on each side of the cylinder and a small satellite. The small satellite is unsteady and lifts off immediately and the two parts finally rebound at $ t/\tau  = 1.79 $. The impact process is in very good agreement with the experiment phenomena shown in Fig. \ref{FIG8} and a video for this case is provided in the supplementary movie 2. For the medium cylinder at $ R/{R_0} = 1.0 $, a smaller part of the droplet spreads onto the plat surface and  the droplet cannot be split into two parts despite the significant deformation. Shortly after the completion of axial retraction, the droplet bounces. Moreover, for the large cylinder radius at $ R/{R_0} = 2.3 $, the droplet film changes to be approximately elliptical in the early stage due to the momentum imbalance. More specifically, more momentum is transferred in the azimuthal direction than the axial direction, as analyzed by Liu \textit{et al.} \cite{liu2015symmetry}. Then the drop retracts first in the axial direction while keeping extending around the cylinder. Once the axial retraction is complete, the droplet bounces. The dynamic process is consistent with the experiment phenomena shown in Fig. \ref{FIG8}. It should be noted that the difference between the last snapshots in Fig. \ref{FIG7}a and Fig. \ref{FIG8}a is because the two snapshots are taken at different instants in the evolution.

\begin{figure}
	\center {
	\includegraphics[width=0.48\textwidth]{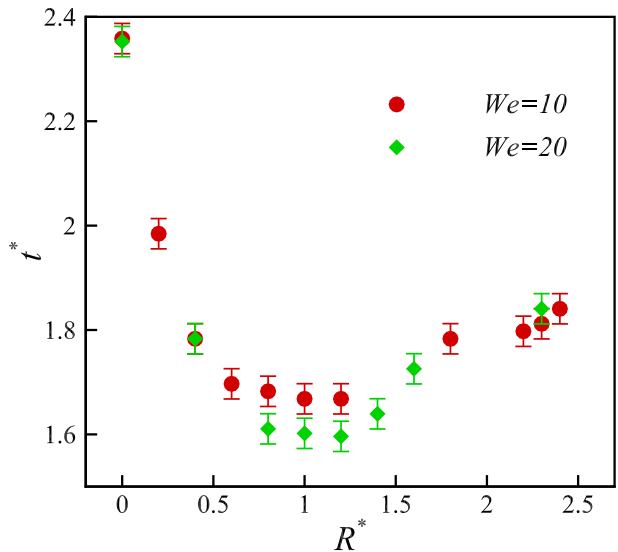}\\
	}
	\caption{\label{FIG9} Variation of the dimensionless contact time
		$ {t^*} = {t_c}/\tau $  as a function of the dimensionless cylinder radius $ {R^*} = R/{R_0} $.
	}
\end{figure}

Figure. \ref{FIG9} presents the variation of the dimensionless contact time
$ {t^*} = {t_c}/\tau $ 
 as a function of the dimensionless cylinder radius $ {R^*} = R/{R_0} $ at  $ We = 10 $ and $ 20 $. Generally, it can be seen that the contact time is minimized at $ {R^*} \approx 1 $. 
Moreover, the figure is 
asymmetric, where the contact time decreases quickly from $ {R^*} = 0 $ to
$ {R^*} \approx 1 $ while it increases slowly at $ {R^*} > 1 $. The trend predicted by the present simulation agrees well with previous studies \cite{liu2015symmetry,andrew2017variation}. In addition, it may be noted that the contact time  for droplet hitting small wires ($ {R^*} < 0.1 $) may be reduced by quite different physical mechanisms and this range is not covered in Fig. \ref{FIG9}. For example, Gauthier \textit{et al.} \cite{gauthier2015water} showed that the contact time for a droplet impact on a small wire scales inversely as the square root of the number of lobes produced.

\begin{figure}
	\center {
		\includegraphics[width=0.48\textwidth]{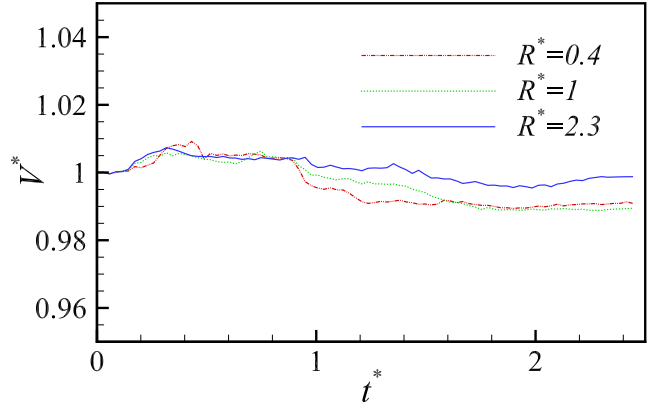}\\
	}
	\caption{\label{fig11} Evolution of dimensionless volume  for the liquid phase  for three cases at $ We=10 $.
	}
\end{figure}

Finally, the conservation of volume or mass in our simulations is considered, which is an important factor affecting the accuracy of simulations involving breakup and/or merging of the phase interfaces \cite{Aniszewski2014}. In Fig.~\ref{fig11}, we show the evolution of dimensionless volume for the liquid phase in the three cases considered in Fig. \ref{FIG7}.
It is seen that the three cases show similar tendency: the volume increases slightly in the early stage and then decreases gradually to a steady value and the fluctuation is approximately within $ \pm 1\% $. This is within the error margin of estimating the volume of complex shapes. In our simulations, we also find that the small secondary droplet produced in Fig.~\ref{FIG7}a is going to be smaller and smaller, and disappear in the end, which is similar to the interface diffusion phenomenon mentioned in \cite{Gongwei2018} and leads to slight loss of liquid volume. Generally, it is demonstrated that the present model performs well in terms of global volume conservation.

\section{Conclusions}\label{sec.5}
Through theoretical analysis, a generalized non-orthogonal multiple-relaxation-time lattice Boltzmann method (MRT-LBM) based on an improved moment set \cite{fei2018three} is developed and proved to reproduce the macroscopic Navier-Stokes equations in the low Mach number limit. 
Compared with the classical MRT-LBM, the numbers of non-zeros in the present transformation matrix ${\bf{M}}$ and its inverse matrix ${{\bf{M}}^{{\rm{ - }}1}}$
are much reduced, leading to a simplified implementation and an enhanced computational efficiency in the present method. Moreover, a non-orthogonal MRT-LBM based on a sub-lattice (e.g., D3Q19) can be extracted from the method on the full-lattice (D3Q27) directly, which shows that the proposed non-orthogonal MRT-LBM exhibits good portability across lattices.

The proposed method is further extended to simulate multiphase flows with large density ratio and tunable surface tension, and
validated through benchmark cases. It is finally applied to two practical problems, a fuel droplet impacting on a dry surface at high Reynolds and Weber numbers and a droplet impacting on a super-hydrophobic wall with a cylindrical obstacle, achieving satisfactory agreement with recent experimental and numerical data.

\section *{Supplementary Material}

See supplementary material for the explicit expressions of
$ {\bf{M}} $ and $ {{\bf{M}}^{ - 1}} $ and supplementary movies.

\section *{Acknowledgments}
The research has received
funding from the MOST National Key Research and Development Programme (Project No. 2016YFB0600805). Further support is gratefully acknowledged from the China Scholarship Council (CSC, No. 201706210262), the European Research Council under
the European Union’s Horizon 2020 Framework Programme(No. FP/2014-2020)/ERC Grant Agreement
No. 739964 ("COPMAT") and the UK Engineering and Physical Sciences Research Council (EPSRC) under the project ``UK Consortium on Mesoscale Engineering Sciences (UKCOMES)" (Grant Nos. EP/L00030X/1 and
EP/R029598/1). QW and KHL would like to gratefully acknowledge the support from The Royal Society$ - $The Natural Science Foundation of China International Exchanges Scheme
(Grant Nos. IE150647 and 51611130192).

\appendix
\section{Chapman-Enskog analysis}\label{sec.5a}
Using Eq. (\ref{e4}) and the relation ${f_i}{\rm{ = }}{{\bar f}_i}{\rm{ + }}\Delta t{{\bar F}_i}/2$, a second-order Taylor series expansion of Eq. (\ref{e9}) at $ ({\bf{x}},t) $ yields
\begin{equation}\label{ea1}
\Delta t({\partial _t} + {{\bf{e}}_i}\cdot\nabla ){f_i} + \frac{{\Delta {t^2}}}{2}{({\partial _t} + {{\bf{e}}_i}\cdot\nabla )^2}{f_i} + O(\Delta {t^3}) =  - {\Lambda _{i,k}}({f_k} - f_k^{eq}) + \Delta t{{\bar F}_i} + \frac{{\Delta {t^2}}}{2}({\partial _t} + {{\bf{e}}_i}\cdot\nabla ){{\bar F}_i}.
\end{equation}
Multiplying Eq. (\ref{ea1}) by the transformation matrix $ {\bf{M}} $  leads to the following equation
\begin{equation}\label{ea2}
({\bf{I}}{\partial _t} + {\bf{D}}){\bf{m}} + \frac{{\Delta t}}{2}{({\bf{I}}{\partial _t} + {\bf{D}})^2}{\bf{m}} + O(\Delta {t^2}) =  - \frac{{\bf{S}}}{{\Delta t}}({\bf{m}} - {{\bf{m}}^{eq}}) + {\bf{\tilde F}} + \frac{{\Delta t}}{2}({\bf{I}}{\partial _t} + {\bf{D}}){\bf{\tilde F}}
\end{equation}
Where $ {\bf{D}} = {{\bf{C}}_x}{\partial _x} + {{\bf{C}}_y}{\partial _y} + {{\bf{C}}_z}{\partial _z} $, in which $ {{\bf{C}}_a} = {\bf{M}}{{\bf{E}}_a}{{\bf{M}}^{ - 1}} $ with $ {{\bf{E}}_a} = diag[{{\bf{e}}_{0,a}},{{\bf{e}}_{1,a}},...,{{\bf{e}}_{26,a}}] $ for $ a = x,y,z  $. By introducing the following Chapman-Enskog expansions,
\begin{equation}\label{ea3}
{\partial _t} = \varepsilon {\partial _{t1}} + {\varepsilon ^2}{\partial _{t2}}{\rm{ + }}...,{\bf{D}}{\rm{ = }}\varepsilon {{\bf{D}}_1},{\bf{m}} = {{\bf{m}}^{eq}} + \varepsilon {{\bf{m}}^{(1)}} + {\varepsilon ^2}{{\bf{m}}^{(2)}} + ...,{\bf{\tilde F}} = \varepsilon {{\bf{\tilde F}}^{(1)}},
\end{equation} 
Eq. (\ref{ea2}) can be rewritten in the consecutive orders of  the expansion parameter $ \varepsilon $  as follows:
\begin{subequations}\label{ea4}
\begin{equation}\label{ea4a}
O(\varepsilon ):({\bf{I}}{\partial _{t1}} + {{\bf{D}}_1}){{\bf{m}}^{eq}} =  - \frac{{\bf{S}}}{{\Delta t}}{{\bf{m}}^{(1)}} + {{\bf{\tilde F}}^{(1)}},
\end{equation}
\begin{equation}\label{ea4b}
O({\varepsilon ^2}):{\partial _{t2}}{{\bf{m}}^{eq}} + ({\bf{I}}{\partial _{t1}} + {{\bf{D}}_1}){{\bf{m}}^{(1)}} + \frac{{\Delta t}}{2}{({\bf{I}}{\partial _{t1}} + {{\bf{D}}_1})^2}{{\bf{m}}^{eq}} =  - \frac{{\bf{S}}}{{\Delta t}}{{\bf{m}}^{(2)}} + \frac{{\Delta t}}{2}({\bf{I}}{\partial _{t1}} + {{\bf{D}}_1}){{{\bf{\tilde F}}}^{(1)}}.
\end{equation}
\end{subequations}
Using the first-order $ O(\varepsilon ) $ equation, the second-order $ O({\varepsilon ^2}) $ can be simplified as 
\begin{equation}\label{ea5}
O({\varepsilon ^2}):{\partial _{t2}}{{\bf{m}}^{eq}} + ({\bf{I}}{\partial _{t1}} + {{\bf{D}}_1})({\bf{I}} - \frac{{\bf{S}}}{2}){{\bf{m}}^{(1)}} =  - \frac{{\bf{S}}}{{\Delta t}}{{\bf{m}}^{(2)}}.
\end{equation}
According to Eq. (\ref{ea4a}) and the definition in Eq. (\ref{e13}), we can obtain continuity and momentum equations at $ O(\varepsilon ) $ level,
\begin{equation}\label{ea6}
\begin{array}{*{20}{c}}
{{\partial _{t1}}\rho  + {\partial _{x1}}(\rho {u_x}) + {\partial _{y1}}(\rho {u_y}) + {\partial _{z1}}(\rho {u_z}) = 0,}  \\
{{\partial _{t1}}(\rho {u_x}) + {\partial _{x1}}(\rho c_s^2 + \rho u_x^2) + {\partial _{y1}}(\rho {u_x}{u_y}) + {\partial _{z1}}(\rho {u_x}{u_z}) = F_x^{(1)},}  \\
{{\partial _{t1}}(\rho {u_y}) + {\partial _{x1}}(\rho {u_x}{u_y}) + {\partial _{y1}}(\rho c_s^2 + \rho u_y^2) + {\partial _{z1}}(\rho {u_y}{u_z}) = F_y^{(1)},}  \\
{{\partial _{t1}}(\rho {u_z}) + {\partial _{x1}}(\rho {u_x}{u_z}) + {\partial _{y1}}(\rho {u_y}{u_z}) + {\partial _{z1}}(\rho c_s^2 + \rho u_z^2) = F_z^{(1)},}  \\
\end{array}
\end{equation}
Analogously, the continuity and $ x$ direction  momentum equations at $ O({\varepsilon ^2}) $ level can be obtained from Eq. (\ref{ea5})
\begin{subequations}\label{ea7}
\begin{equation}\label{ea7a}
{\partial _{t2}}\rho  = 0,
\end{equation}
\begin{equation}\label{ea7b}
{\partial _{t2}}(\rho {u_x}) + \frac{1}{3}{\partial _{x1}}\left[ {(1 - \frac{{{s_b}}}{2})m_7^{(1)} + (1 - \frac{{{s_\nu }}}{2})m_8^{(1)} + (1 - \frac{{{s_\nu }}}{2})m_9^{(1)}} \right] + {\partial _{y1}}\left[ {(1 - \frac{{{s_\nu }}}{2})m_4^{(1)}} \right] + {\partial _{z1}}\left[ {(1 - \frac{{{s_\nu }}}{2})m_5^{(1)}} \right] = 0,
\end{equation}
\end{subequations}
where the unknown first-order non-equilibrium moments can be obtained according to Eq. (\ref{ea4a}),
\begin{equation}\label{ea8}
\begin{array}{*{20}{c}}
{\partial _{t1}}m_4^{eq} + {\partial _{x1}}m_{12}^{eq} + {\partial _{y1}}m_{10}^{eq} + {\partial _{z1}}m_{16}^{eq} =  - {s_\nu }m_4^{(1)}/\Delta t + \tilde F_4^{(1)}, \\
{{\partial _{t1}}m_5^{eq} + {\partial _{x1}}m_{13}^{eq} + {\partial _{y1}}m_{16}^{eq} + {\partial _{z1}}m_{11}^{eq} =  - {s_\nu }m_5^{(1)}/\Delta t + \tilde F_5^{(1)},}  \\
{{\partial _{t1}}m_7^{eq} + {\partial _{x1}}(m_1^{eq} + m_{10}^{eq} + m_{11}^{eq}) + {\partial _{y1}}(m_2^{eq} + m_{12}^{eq} + m_{14}^{eq}) + {\partial _{z1}}(m_3^{eq} + m_{13}^{eq}{\rm{ + }}m_{15}^{eq}) =  - {s_b}m_7^{(1)}/\Delta t + \tilde F_7^{(1)},}  \\
{{\partial _{t1}}m_8^{eq} + {\partial _{x1}}(m_1^{eq} - m_{10}^{eq}) + {\partial _{y1}}( - m_2^{eq} + m_{12}^{eq}) + {\partial _{z1}}(m_{13}^{eq} - m_{15}^{eq}) =  - {s_\nu }m_8^{(1)}/\Delta t + \tilde F_8^{(1)},}  \\
{{\partial _{t1}}m_9^{eq} + {\partial _{x1}}(m_1^{eq} - m_{11}^{eq}) + {\partial _{y1}}(m_{12}^{eq} - m_{14}^{eq}) + {\partial _{z1}}( - m_3^{eq} + m_{13}^{eq}) =  - {s_\nu }m_9^{(1)}/\Delta t + \tilde F_9^{(1)},}  \\
\end{array}
\end{equation}
Here, the following relations can be obtained according to Eq. (\ref{ea6}),
\begin{equation}\label{ea9}
{\partial _{t1}}(\rho {u_\alpha }{u_\beta }) = {u_\alpha }{\partial _{t1}}(\rho {u_\beta }) + {u_\beta }{\partial _{t1}}(\rho {u_\alpha }) - {u_\alpha }{u_\beta }{\partial _{t1}}(\rho ) \approx  - {u_\alpha }{\partial _{\beta 1}}(\rho c_s^2) - {u_\beta }{\partial _{a1}}(\rho c_s^2) + {u_\alpha }F_\beta ^{(1)} + {u_\beta }F_\alpha ^{(1)}
\end{equation}
Substituting the above relations into Eq. (\ref{ea8}), the first-order non-equilibrium moments in Eq. (\ref{ea7b}) can be written explicitly,
\begin{equation}\label{ea10}
\begin{array}{l}
m_4^{(1)} =  - \frac{{\Delta t}}{{{s_\nu }}}\rho c_s^2({\partial _{x1}}{u_y} + {\partial _{y1}}{u_x}),~m_5^{(1)} =  - \frac{{\Delta t}}{{{s_\nu }}}\rho c_s^2({\partial _{x1}}{u_z} + {\partial _{z1}}{u_x}),~m_7^{(1)} =  - \frac{{2\Delta t}}{{{s_b}}}\rho c_s^2({\partial _{x1}}{u_x} + {\partial _{y1}}{u_y} + {\partial _{z1}}{u_z}), \\ 
m_8^{(1)} =  - \frac{{2\Delta t}}{{{s_\nu }}}\rho c_s^2({\partial _{x1}}{u_x} - {\partial _{y1}}{u_y}),~m_9^{(1)} =  - \frac{{2\Delta t}}{{{s_\nu }}}\rho c_s^2({\partial _{x1}}{u_x} - {\partial _{z1}}{u_z}). \\ 
\end{array}
\end{equation}
Thus Eq. (\ref{ea7b}) can be rewritten as
\begin{equation}\label{ea11}
\begin{array}{l}
{\partial _{t2}}(\rho {u_x}) = {\partial _{x1}}\left[ {\rho {\nu _b}({\nabla _1}\cdot{\bf{u}}) + \frac{2}{3}\rho \nu (2{\partial _{x1}}{u_x} - {\partial _{y1}}{u_y} - {\partial _{z1}}{u_z})} \right] \\ 
+ {\partial _{y1}}\left[ {\rho \nu ({\partial _{y1}}{u_x} + {\partial _{x1}}{u_y})} \right] + {\partial _{z1}}\left[ {\rho \nu ({\partial _{z1}}{u_x} + {\partial _{x1}}{u_z})} \right], \\ 
\end{array}
\end{equation}
where the kinematic viscosity and bulk viscosity are given by
\begin{equation}\label{ea12}
\nu  = c_s^2(\frac{1}{{{s_\nu }}} - \frac{1}{2})\Delta t,~~{\nu _b} = \frac{2}{3}c_s^2(\frac{1}{{{s_b}}} - \frac{1}{2})\Delta t.
\end{equation}
Similarly, the momentum equations in $ y $ and $ z $ directions at   $ O({\varepsilon ^2}) $ level are given as
\begin{equation}\label{ea13}
\begin{array}{l}
{\partial _{t2}}(\rho {u_y}) = {\partial _{x1}}\left[ {\rho \nu ({\partial _{y1}}{u_x} + {\partial _{x1}}{u_y})} \right] + {\partial _{y1}}\left[ {\rho {\nu _b}({\nabla _1}\cdot{\bf{u}}) + \frac{2}{3}\rho \nu (2{\partial _{y1}}{u_y} - {\partial _{x1}}{u_x} - {\partial _{z1}}{u_z})} \right] \\ 
+ {\partial _{z1}}\left[ {\rho \nu ({\partial _{z1}}{u_y} + {\partial _{y1}}{u_z})} \right] \\ 
{\partial _{t2}}(\rho {u_z}) = {\partial _{x1}}\left[ {\rho \nu ({\partial _{z1}}{u_x} + {\partial _{x1}}{u_z})} \right] + {\partial _{y1}}\left[ {\rho \nu ({\partial _{z1}}{u_y} + {\partial _{y1}}{u_z})} \right] \\ 
+ {\partial _{z1}}\left[ {\rho {\nu _b}({\nabla _1}\cdot{\bf{u}}) + \frac{2}{3}\rho \nu (2{\partial _{z1}}{u_z} - {\partial _{x1}}{u_x} - {\partial _{y1}}{u_y})} \right]. \\ 
\end{array}
\end{equation}
Combining the $ O({\varepsilon}) $ level and $ O({\varepsilon ^2}) $ level equations using the expansion relations in Eq. (\ref{ea3}), we can obtain,
\begin{equation}\label{ea14}
\begin{array}{l}
{\partial _t}\rho  + \nabla  \cdot (\rho {\bf{u}}) = 0, \\ 
{\partial _t}(\rho {\bf{u}}) + \nabla  \cdot (\rho {\bf{uu}}) =  - \nabla (\rho c_s^2) + \nabla  \cdot \left[ {\rho \nu (\nabla {\bf{u}} + {{(\nabla {\bf{u}})}^{\rm T}}) - \frac{2}{3}\rho \nu (\nabla  \cdot {\bf{u}}){\bf{I}}} \right] + \nabla \left[ {\rho {\nu _b}(\nabla  \cdot {\bf{u}})} \right] + {\bf{F}}. \\ 
\end{array}
\end{equation}
From the above Chapman-Enskog analysis, we can see that the Navier-Stokes equations can be correctly recovered from the present non-orthogonal MRT-LB model in the low Mach number limit.

\section{Pseudopotential Multiphase model with high density ratio and tunable surface tension}\label{sec.5a2}
When the square-root-form pseudopotential is used, the mechanical stability condition can not be accurately satisfied in the pseudopotential model. To solve this problem, Li \textit{et al.}  proposed a modified forcing scheme to adjust the mechanical stability condition \cite{li2012forcing,li2013lattice}. In addition, the original Shan-Chen pseudopotential model also suffers from the problem that the surface tension cannot be tuned independently of the density ratio. Li \textit{et al.}
developed another method to tune the surface tension in the 2D MRT LBM by incorporating a source term into the LB equation \cite{li2013achieving}. Due to the simplicity and efficiency, the above mentioned methods \cite{li2012forcing,li2013lattice,li2013achieving} have been adopted by different researchers \cite{xu2015three,gong2017thermal,xu2017lattice}.
Inspired by Li \textit{et al.} \cite{li2012forcing,li2013lattice,li2013achieving}, to achieve large density ratio and tunable surface tension in the present non-orthogonal MRT-LBM, several elements in  $ {\bf{\tilde F}} $ can be modified as,

\begin{equation}\label{e17}
\begin{array}{l}
\tilde F_4^{'} = {{\tilde F}_4} - \frac{{{Q_{xy}}}}{{(s_\nu ^{ - 1} - 0.5)\Delta t}},~~\tilde F_5^{'} = {{\tilde F}_5} - \frac{{{Q_{xz}}}}{{(s_\nu ^{ - 1} - 0.5)\Delta t}},~~\tilde F_6^{'} = {{\tilde F}_6} - \frac{{{Q_{yz}}}}{{(s_\nu ^{ - 1} - 0.5)\Delta t}}, \\ 
\tilde F_7^{'} = {{\tilde F}_7} + \frac{{6\sigma {{\left| {{{\bf{F}}_{{\mathop{\rm int}} }}} \right|}^2}}}{{{\psi ^2}(s_e^{ - 1} - 0.5)\Delta t}} + \frac{{4({Q_{xx}} + {Q_{yy}} + {Q_{zz}})}}{{5(s_e^{ - 1} - 0.5)\Delta t}},~~\tilde F_8^{'} = {{\tilde F}_8} - \frac{{({Q_{xx}} - {Q_{yy}})}}{{(s_\nu ^{ - 1} - 0.5)\Delta t}},~~\tilde F_9^{'} = {{\tilde F}_9} - \frac{{({Q_{xx}} - {Q_{zz}})}}{{(s_\nu ^{ - 1} - 0.5)\Delta t}}, \\ 
\end{array}
\end{equation}
where the parameter $ \sigma $, usually within $ 0.0625 \le \sigma  \le 0.125 $,  is employed to adjust the mechanical stability condition and its exact value can be determined by fitting the liquid-gas coexistence densities. The variable $ {Q_{\alpha \beta }} $ is obtained via \cite{li2013achieving}, 
\begin{equation}\label{e18}
{Q_{\alpha \beta }} = \kappa \frac{G}{2}\psi ({\bf{x}})\sum\limits_i {w({{\left| {{{\bf{e}}_i}} \right|}^2})} \left[ {\psi ({\bf{x}} + {{\bf{e}}_i}\Delta t) - \psi ({\bf{x}})} \right]{{\rm{e}}_{i\alpha }}{e_{i\beta }},
\end{equation}
where the parameter $ \kappa $  is used to tune the surface tension. Consistent with Eq. (16), only the nearest or single-range neighboring nodes are needed in the calculation of  $ {Q_{\alpha \beta }} $, thus no additional computational complexity is introduced even at the boundary nodes. 

In the Chapman analysis, the above modifications do not affect the equations at  $ O({\varepsilon}) $  level. For the equations at  $ O({\varepsilon ^2}) $  level, it can be seen that Eq. (\ref{ea10}) is changed due to the modifications of in the forcing terms. As a result, Eq. (\ref{ea11}) is changed correspondingly,
\begin{equation}\label{ea15}
\begin{array}{l}
{\partial _{t2}}(\rho {u_x}) = {\partial _{x1}}\left[ {\rho {\nu _b}({\nabla _1} \cdot {\bf{u}}) + \frac{2}{3}\rho \nu (2{\partial _{x1}}{u_x} - {\partial _{y1}}{u_y} - {\partial _{z1}}{u_z}) - \frac{{2\sigma {{\left| {{\bf{F}}_{{\mathop{\rm int}} }^{(1)}} \right|}^2}}}{{{\psi ^2}}} + (\frac{2}{5}{Q_{xx1}} - \frac{3}{5}{Q_{yy1}} - \frac{3}{5}{Q_{zz1}})} \right] \\ 
{\rm{              }} + {\partial _{y1}}\left[ {\rho \nu ({\partial _{y1}}{u_x} + {\partial _{x1}}{u_y}) + {Q_{xy1}}} \right] + {\partial _{z1}}\left[ {\rho \nu ({\partial _{z1}}{u_x} + {\partial _{x1}}{u_z}) + {Q_{xz1}}} \right]. \\ 
\end{array}
\end{equation}
The Taylor expansions of the interaction force  $ {{\bf{F}}_{{\mathop{\rm int}} }} $ and the term  $ {\bf{Q}} $ yield \cite{li2012forcing,li2013achieving,xu2015three}
\begin{equation}\label{ea16}
{{\bf{F}}_{{\mathop{\rm int}} }} =  - G{c^2}\left[ {\psi \nabla \psi  + \frac{1}{6}{c^2}\psi \nabla ({\nabla ^2}\psi ) + ...} \right],~~{\bf{Q}} = \frac{1}{{12}}\kappa G{c^4}\left[ {\psi {\nabla ^2}\psi {\bf{I}} + 2\psi \nabla \nabla \psi  + ...} \right]
\end{equation}
Substituting Eq. (\ref{ea16}) into Eq. (\ref{ea15}) and combining the correspondingly modified equations in \textit{y} and \textit{z} directions, the macroscopic momentum equation in Eq. (\ref{ea15}) is rewritten as,
\begin{equation}\label{ea17}
\begin{array}{*{20}{c}}
{{\partial _t}(\rho {\bf{u}}) + \nabla \cdot(\rho {\bf{uu}}) =  - \nabla (\rho c_s^2) + \nabla \cdot\left[ {\rho \nu (\nabla {\bf{u}} + {{(\nabla {\bf{u}})}^{\rm{T}}}) - \frac{2}{3}\rho \nu (\nabla \cdot{\bf{u}}){\bf{I}}} \right] + \nabla \left[ {\rho {\nu _b}(\nabla \cdot{\bf{u}})} \right] + {\bf{F}}}  \\
{ - 2G^2{c^4}\sigma \nabla \cdot\left( {{{\left| {\nabla \psi } \right|}^2}{\bf{I}}} \right) - \nabla \cdot\left[ {\kappa \frac{{G{c^4}}}{6}(\psi {\nabla ^2}\psi {\bf{I}} - \psi \nabla \nabla \psi )} \right].}  \\
\end{array}
\end{equation}
Following the standard approach by Shan \cite{shan2008pressure}, it can be shown that the surface tension coefficient finally reads \cite{li2013achieving,xu2015three}
\begin{equation}\label{ea18}
\gamma  = \int_{ - \infty }^\infty  {({p_0} - {p_T})dx}  =  - \frac{{G{c^4}(1 - \kappa )}}{6}\int_{{\rho _g}}^{{\rho _l}} {\psi {'^2}(\frac{{d\rho }}{{dx}})d} \rho 
\end{equation}
where the integral extends across a  planar interface normal to the $ x $ direction, $ {p_0} $  is the normal pressure tensor, $ {p_T} $  is the transversal pressure tensor, and  $ \psi ' = d\psi /d\rho $ . From the above, it is shown that the surface tension  $ \gamma $ is proportional to  $ (1 - \kappa ) $ and can be tuned independently of the density ratio.
\\

\section{D3Q19 non-orthogonal MRT-LBM}\label{sec.5b}
For the D3Q19 lattice, the discrete velocities $ {{\bf{e}}_j} = [\left| {{e_{jx}}} \right\rangle ,\left| {{e_{jy}}} \right\rangle ,\left| {{e_{jz}}} \right\rangle ] $  ($ j = 0,1,..,18 $) are the first 19 elements in the D3Q27 lattice,
\begin{equation}
\begin{array}{l}
\left| {{e_{jx}}} \right\rangle  = {\left[ {0,{\rm{ 1}}, - {\rm{1}},{\rm{ }}0,{\rm{ }}0,{\rm{ }}0,{\rm{ }}0,{\rm{ 1}}, - {\rm{1}},{\rm{ 1}}, - {\rm{1}},{\rm{ 1}}, - {\rm{1}},{\rm{ 1}}, - {\rm{1}},{\rm{ }}0,{\rm{ }}0,{\rm{ }}0,{\rm{ }}0} \right]^{\rm T}}, \\ 
\left| {{e_{jx}}} \right\rangle  = {\left[ {0,{\rm{ }}0,{\rm{ }}0,{\rm{ 1}}, - {\rm{1}},{\rm{ }}0,{\rm{ }}0,{\rm{ 1}},{\rm{ 1}}, - {\rm{1}}, - {\rm{1}},{\rm{ }}0,{\rm{ }}0,{\rm{ }}0,{\rm{ }}0,{\rm{ 1}}, - {\rm{1}},{\rm{ 1}}, - {\rm{1}}} \right]^{\rm T}}, \\ 
\left| {{e_{jz}}} \right\rangle  = {\left[ {0,{\rm{ }}0,{\rm{ }}0,{\rm{ }}0,{\rm{ }}0,{\rm{ 1}}, - {\rm{1}},{\rm{ }}0,{\rm{ }}0,{\rm{ }}0,{\rm{ }}0,{\rm{ 1}},{\rm{ 1}}, - {\rm{1}}, - {\rm{1}},{\rm{ 1}},{\rm{ 1}}, - {\rm{1}}, - {\rm{1}}} \right]^{\rm T}}. \\ 
\end{array}
\end{equation}
The raw moment set $ {\bf{m}} = {[{m_0},{m_1},...,{m_{18}}]^{\rm T}} $ can be extracted from Eq. (\ref{e11}),
\begin{equation}
\begin{array}{l}
{\bf{m}} = [{k_{000}},{k_{100}},{k_{010}},{k_{001}},{k_{110}},{k_{101}},{k_{011}},{k_{200}} + {k_{020}} + {k_{002}},{k_{200}} - {k_{020}}, \\ 
{k_{200}} - {k_{002}},{k_{120}},{k_{102}},{k_{210}},{k_{201}},{k_{012}},{k_{021}},{k_{220}},{k_{202}},{k_{022}}{]^{\rm{T}}} \\ 
\end{array}
\end{equation}
so do the relaxation matrix,
\begin{equation}
{\bf{S}} = diag({s_0},{s_1},{s_1},{s_1},{s_2},{s_2},{s_2},{s_{2b}},{s_2},{s_2},{s_3},{s_3},{s_3},{s_3},{s_3},{s_3},{s_4},{s_4},{s_4}).
\end{equation}
The transformation matrix here is a $ 19 \times 19 $ matrix and is extracted from the corresponding rows and columns in Eq. (\ref{matrix}) directly,
\begin{equation}
\setlength{\arraycolsep}{1.7pt}
{\bf{M} = }\left[ 
\begin{array}{l r r r r r r r r r r r r r r r r r r r r r r r r r r}
1 &1 &1&1&1&1&1&1&1&1 &1 &1&1&1&1&1&1&1&1\\

0 &1 & -1&0&0&0&0&1&-1&1&-1&1&-1&1&-1&0&0&0&0\\
0 &0 &0&1&-1&0&0&1&1&-1 &-1 &0&0&0&0&1&-1&1&-1\\
0 &0 &0&0&0&1&-1&0&0&0&0 &1&1&-1&-1&1&1&-1&-1\\

0 &0 &0&0&0&0&0&1&-1&-1&1&0&0&0&0&0&0&0&0\\
0 &0 &0&0&0&0&0&0&0&0&0&1&-1&-1&1&0&0&0&0\\
0 &0 &0&0&0&0&0&0&0&0&0&0&0&0&0&1&-1&-1&1\\

0 &1 &1&1&1&1&1&2&2&2 &2 &2&2&2&2&2&2&2&2\\
0 &1 &1&-1&-1&0&0&0&0&0 &0 &1&1&1&1&-1&-1&-1&-1\\
0 &1 &1&0&0&-1&-1&1&1&1 &1 &0&0&0&0&-1&-1&-1&-1\\

0 &0 &0&0&0&0&0&1&-1&1 &-1 &0&0&0&0&0&0&0&0\\
0 &0 &0&0&0&0&0&0&0&0 &0 &1&-1&1 &-1&0&0&0&0\\
0 &0 &0&0&0&0&0&1&1&-1 &-1 &0&0&0&0&0&0&0&0\\
0 &0 &0&0&0&0&0&0&0&0 &0 &1&1&-1&-1&0&0&0&0\\
0 &0 &0&0&0&0&0&0&0&0 &0 &0&0&0&0&1&-1&1&-1\\
0 &0 &0&0&0&0&0&0&0&0 &0 &0&0&0&0&1&1&-1&-1\\

0 &0 &0&0&0&0&0&1&1&1 &1 &0&0&0&0&0&0&0&0\\
0 &0 &0&0&0&0&0&0&0&0&0 &1&1&1&1&0&0&0&0\\
0 &0 &0&0&0&0&0&0&0&0&0 &0&0&0&0&1&1&1&1\\
\end{array} \right].
\end{equation}
Its inverse $ {{\bf{M}}^{ - 1}} $ can be easily obtained by software like MATLAB. It is also shown that $ {{\bf{M}}^{ - 1}} $ for the D3Q19 model can be extracted from the corresponding rows and columns for the D3Q27 $ {{\bf{M}}^{ - 1}} $ (see in the Supplementary Material). In the same way, equilibrium raw moments and the forcing terms are given respectively,
\begin{equation}
\begin{array}{l}
{{\bf{m}}^{eq}} = [\rho ,\rho {u_x},\rho {u_y},\rho {u_z},\rho {u_x}{u_y},\rho {u_x}{u_z},\rho {u_y}{u_z},\rho (1 + {{\bf{u}}^2}),\rho (u_x^2 - u_y^2),\rho (u_x^2 - u_z^2),\rho c_s^2{u_x},\rho c_s^2{u_x}, \\ 
\rho c_s^2{u_y},\rho c_s^2{u_z},\rho c_s^2{u_y},\rho c_s^2{u_z},\rho c_s^2(c_s^2 + u_x^2 + u_y^2),\rho c_s^2(c_s^2 + u_x^2 + u_z^2),\rho c_s^2(c_s^2 + u_y^2 + u_z^2){]^{\rm T}}{\rm{ }} \\ 
\end{array}
\end{equation}
\begin{equation}
\begin{array}{l}
{\bf{\tilde F}} = [0,{F_x},{F_y},{F_z},{\rm{ }}{F_x}{u_y} + {F_y}{u_x},{F_x}{u_z} + {F_z}{u_x},{F_y}{u_z} + {F_z}{u_y},2{\bf{F}} \cdot {\bf{u}},2({F_x}{u_x} - {F_y}{u_y}),2({F_x}{u_x} - {F_z}{u_z}), \\ 
{F_x}c_s^2,{F_x}c_s^2,{F_y}c_s^2,{F_z}c_s^2,{F_y}c_s^2,{F_z}c_s^2,2c_s^2({F_x}{u_x} + {F_y}{u_y}),2c_s^2({F_x}{u_x} + {F_z}{u_z}),2c_s^2({F_y}{u_y} + {F_z}{u_z}){]^{\rm T}} \\ 
\end{array}
\end{equation}

Through the Chapman-Enskog analysis, it can be seen that the Navier-Stokes equations can be correctly recovered from the D3Q19 non-orthogonal MRT-LB model in the low Mach number limit. When coupled with the pseudopotential multiphase model, the modification method in Eq. (\ref{e17}) is also applicable. It can be further shown that the D2Q9 non-orthogonal MRT-LBM in \cite{lycett2014multiphase} can be extracted from the present D3Q27 model easily.
From the above, we can see the non-orthogonal MRT-LBM has very good portability across lattices (a model in a sub-lattice can be extracted from the model in the full-lattice directly), while the conventional orthogonal MRT-LBM does not have this feature, to the best of our knowledge.
\section*{References}

%

\end{document}